\documentclass[prb,twocolumn,superscriptaddress,longbibliography]{revtex4-2}
\usepackage{diagbox} 
\usepackage{graphicx}
\usepackage{pstricks}
\usepackage{pst-node}
\usepackage{pstricks-add}
\usepackage{amsfonts,amssymb,amsmath}
\usepackage[colorlinks=true,citecolor=green,linkcolor=red,urlcolor=blue]{hyperref}
\usepackage{subfigure}
\usepackage{calligra}

\usepackage{ulem}

\usepackage{dsfont}

\newcommand{\beq}{\begin{equation}}
\newcommand{\be}{\begin{equation}}
\newcommand{\beqn}{\begin{eqnarray}}
\newcommand{\eeq}{\end{equation}}
\newcommand{\ee}{\end{equation}}
\newcommand{\eeqn}{\end{eqnarray}}
\newcommand{\nn}{\nonumber}

\newcommand{\bem}{\begin{pmatrix}}
\newcommand{\eem}{\end{pmatrix}}

\newlength{\ldag}
\newcommand{\bra}[1]{\langle#1|}
\newcommand{\ket}[1]{|#1\rangle}

\newcommand{\tr}{\mathrm{Tr}}

\begin{document}

\title{Topological and nontopological degeneracies in generalized string-net models}

\author{Anna Ritz-Zwilling}
\email{anna.ritz\_zwilling@sorbonne-universite.fr }
\affiliation{Sorbonne Universit\'e, CNRS, Laboratoire de Physique Th\'eorique de la Mati\`ere Condens\'ee, LPTMC, F-75005 Paris, France}

\author{Jean-No\"el Fuchs}
\email{jean-noel.fuchs@sorbonne-universite.fr}
\affiliation{Sorbonne Universit\'e, CNRS, Laboratoire de Physique Th\'eorique de la Mati\`ere Condens\'ee, LPTMC, F-75005 Paris, France}

\author{Steven H. Simon}
\email{steven.simon@physics.ox.ac.uk}
\affiliation{Rudolf Peierls Centre for Theoretical Physics, Clarendon Laboratory, Oxford, OX1 3PU, United Kingdom}

\author{Julien Vidal}
\email{julien.vidal@sorbonne-universite.fr}
\affiliation{Sorbonne Universit\'e, CNRS, Laboratoire de Physique Th\'eorique de la Mati\`ere Condens\'ee, LPTMC, F-75005 Paris, France}
%%%%%%%%%%%%%%%%%%%%%%%%%%%%%%%%%
\begin{abstract}
Generalized string-net models have been recently proposed in order to enlarge the set of possible topological quantum phases emerging from the original string-net construction. In the present work, we do not consider vertex excitations and restrict to plaquette excitations, or fluxons, that satisfy important identities. We explain how to compute the energy-level degeneracies of the generalized string-net Hamiltonian associated to an arbitrary unitary fusion category. 
%, including those with noncommutative fusion rules
In contrast to the degeneracy of the ground state, which is purely topological, that of excited energy levels depends not only on the Drinfeld center of the category, but also on internal multiplicities obtained from the tube algebra defined from the category. For a noncommutative category, these internal multiplicities result in extra nontopological degeneracies. Our results are valid for any trivalent graph and any orientable surface. We illustrate our findings with nontrivial examples.
%OLD VERSION Generalized string-net models have been recently proposed in order to enlarge the set of possible topological quantum phases emerging from the original string-net construction. In the present work, we explain how to compute the energy-level degeneracies of the generalized string-net Hamiltonian associated to an arbitrary unitary fusion category, including those with noncommutative fusion rules. These degeneracies are expressed in terms of the quantum dimensions  of the simple objects in the Drinfeld center and, in the noncommutative case, by their internal multiplicities, defined from a tube algebra. Our results are valid for any trivalent graph and any orientable surface.%
\end{abstract}
%%%%%%%%%%%%%%%%%%%%%%%%%%%%%%%%%
%\date{}

\maketitle

\tableofcontents

%%%%%%%%%%%%%%%%%%%%%%%%%%%%%%%%%
%%%%%%%%%%%%%%%%%%%%%%%%%%%%%%%%%
%%%%%%%%%%%%%%%%%%%%%%%%%%%%%%%%%
%%%%%%%%%%%%%%%%%%%%%%%%%%%%%%%%%

%
%
%%%%%%%%%%%%%%%%%%%
%%%%%%%%%%%%%%%%%%%
\section{Introduction}
\label{sec:intro}
%%%%%%%%%%%%%%%%%%%
%%%%%%%%%%%%%%%%%%%
%
%
The idea that two-dimensional quantum systems could contain quasiparticles with exchange statistics different from bosons and fermions emerged about fifty years ago~\cite{Leinaas77,Wilczek82}. These point-like particles, called $\textit{anyons}$, could,  under mutual exchange, accumulate ``any", possibly fractional, phase (Abelian anyons), or even change into a new state orthogonal to the initial one (non-Abelian anyons).
%The idea of anyons, quasiparticles with exchange statistics different from bosons or fermions, emerged about fifty years ago ~\cite{Wilczek82, Leinaas77}. It is in fact possible in two dimensions to have point-like particles which, under mutual exchange, accumulate ``any", possibly fractional, phase (Abelian anyons), or even change into a new state orthogonal to the initial one (non-Abelian anyons). 
Yet, only recent experiments provided evidence for the presence of Abelian anyons in the fractional quantum Hall effect~\cite{Bartolomei20,Nakamura20}, and investigated non-Abelian anyons in quantum processors made of superconducting qubits~\cite{GoogleAI22,Xu23} or trapped ions~\cite{Iqbal23}. Anyons are a fundamental characteristic of topologically-ordered phases: gapped phases that appear in strongly interacting or frustrated systems and that cannot be identified by local symmetries. In particular, anyons are connected to one of
%topological order's most typical feature:
the defining properties of topological order: a ground-state degeneracy  which depends on the surface topology on which the system is defined, and which is robust against small local perturbations (a fact which motivated the idea of using topological order to implement fault tolerant quantum computation ~\cite{Kitaev03,Nayak08}). The excited states of topologically-ordered phases also show nontrivial degeneracies related to the fusion of anyons.

Some of the most studied models for topologically-ordered phases are the {\it string-net} models introduced by Levin and Wen~\cite{Levin05}, and their generalizations~\cite{Lin14,Lan14,Hahn20,Lin21}. These models, while not being able to produce all possible topological orders, are believed to produce exactly those bosonic topological orders that have gappable edges (and are therefore achiral). The string-net models are exactly solvable. While the energy levels are easy to obtain, the corresponding spectrum degeneracies are nontrivial and are the main focus of the present paper.

\begin{figure}[h]
\includegraphics[width=0.65\columnwidth]{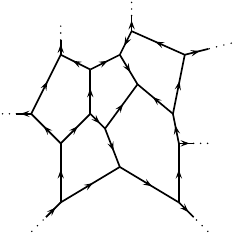}
\caption{A portion of a trivalent graph with oriented edges.}
\label{fig:trivalentlattice}
\end{figure}

The string-net models are built from planar diagram algebras --- so-called (spherical) unitary fusion categories (UFCs).   A UFC is equipped with a set of quantum numbers (``particles" or ``simple objects" or ``labels'' or ``strings'') obeying certain fusion rules.   The string-net model is then built on some trivalent graph [dual to a triangulation of an orientable two-dimensional  (2D) manifold] as shown in Fig.~\ref{fig:trivalentlattice}.  The full Hilbert space is given by assigning a label (a simple object) to each directed edge of the graph.   
%
%
%%%%%%%%%%%%
%%%%%%%%%%%%
%
%
Given a UFC $\mathcal{C}$, the string-net model built from $\mathcal{C}$ realizes a topological order~\cite{Kirillov11} corresponding to the Drinfeld center $\mathcal{Z}(\mathcal{C})$~\cite{Wang_book} which  is a well-defined (2+1)D topological quantum field theory (TQFT), or, in mathematical terms, a unitary modular tensor category (UMTC). For example, the vector space dimension assigned to a 2D compact manifold of genus $g$ by $\mathcal{Z}(\mathcal{C})$ matches the ground-state degeneracy of the string-net model built from $\mathcal{C}$ on the same manifold.   However, what is often not appreciated is that the agreement between the vector space dimension assigned by $\mathcal{Z}({\mathcal{C}})$ and the degeneracy of states in the string-net model may not be perfect when one considers string-nets on manifolds with boundary, or the case where quasiparticles are present, i.e., excitations above the ground state.  In particular, for cases other than the ground state on a closed manifold,  the string-net model displays both topological and nontopological degeneracies.

 By ``topological degeneracies," we mean degeneracies that arise from multiple topological fusion channels in the ${\cal Z}({\cal C})$ topological quantum field theory.  As we will discuss below in section  \ref{sec:spdeg}, the degeneracy of the ground state on a surface of nonzero genus is such a topological degeneracy.   This type of degeneracy cannot be split by local perturbations so long as the excitation gap does not close.    Here, the notion of locality is related to  operators acting on a typical scale smaller than the systole (shortest noncontractible loops).    Degeneracies that arise from multiple fusion channels of quasiparticle excitations, if present, will also be called topological.  These can be split by perturbations only if the perturbing operators connect the positions of the multiple excitations.   So for example, if two excitations are very close to each other, a fairly local perturbation operator could in principle split such degeneracy.   Topological degeneracies, independent of how easy they are to split, are encoded in the fusion rules of $\mathcal{Z}(\mathcal{C})$.

However, in the case of a string-net model built from a noncommutative UFC there are additional degeneracies associated with quasiparticles that are not topological,  and counting them requires the knowledge of other quantities that will be discussed below.    Indeed, in a given topological sector, one may still have some additional degeneracies which can be lifted by completely local perturbations.  

For a string-net model built from a category $\cal C$, certain quasiparticle types of the emergent theory  $\mathcal{Z}(\mathcal{C})$ play a special role and are denoted as ``fluxons" or plaquette excitations. These elementary excitations are always bosons that can be excited in the string-net model without violating the fusion condition of the model at every trivalent vertex (see Sec.~\ref{sec:gsnm} for more details). In the current paper,  we will focus on this type of excitations. 

Excitations that are not fluxons and boundaries that are not smooth generally require violations of the fusion condition. This is more complicated since there may be  inequivalent ways to define the model once the fusion conditions are violated.   Note that in Ref.~\cite{Hu18} a particular extension of the string-net model (known as ``extended string-net model'') is considered that contains all topological quasiparticle types but never violates any vertex constraints at the price of some other complications.   We leave for future work analysis of that extended model. 

It is worth noting that being a fluxon for some particular particle types of a UMTC is dependent on the microscopic string-net model and is not a property of the TQFT.   In particular it may happen that the same TQFT may be constructed either as the Drinfeld center   $\mathcal{Z}(\mathcal{C})$ of a UFC $\mathcal{C}$ or as the Drinfeld center   $\mathcal{Z}(\mathcal{C}')$ of another UFC $\mathcal{C}'$. When $\mathcal{Z}(\mathcal{C}') \simeq \mathcal{Z}(\mathcal{C}) $  we say that $\mathcal{C}$ and $\mathcal{C}'$ are {\it Morita equivalent}~\cite{NoteMorita}.  However, the fluxons of the string-net model built from $\mathcal{C}$ are generally different from the fluxons of the string-net model built from $\mathcal{C}'$.  This is quite natural since a fluxon is defined as satisfying the fusion rules of the underlying UFC that the string-net model is built from.

In the case where the fusion rules of the category $\mathcal C$ are commutative, i.e., when $a\times b = b\times a$ for all objects $a$ and $b$ in the category, there is a substantial simplification since there are no internal multiplicities.  One can then find the topological degeneracy of a 2D orientable manifold  with any number of quasiparticles on it using classic formulas found by Moore, Seiberg and Banks~\cite{Moore89} and Verlinde~\cite{Verlinde88}.   
However, in the case where the fusion rules of the category $\mathcal C$ are noncommutative, i.e., $a\times b \neq b\times a$ for some $a$ and $b$ in the category, the situation is more complicated.  Here, for some of the fluxons, there is an additional internal degeneracy factor on top of the TQFT/conformal field theory prediction of  Moore, Seiberg,  Banks, and Verlinde~\cite{Moore89,Verlinde88}. We will be  particularly interested in this case. 

It may seem to be quite an unusual situation to have a fusion category with noncommutative fusion rules.   However, this is precisely the case for the famous Kitaev quantum double model~\cite{Kitaev03}  for a non-Abelian group, which is the first (and potentially simplest) anyon model known to be universal for quantum computation. Here, we are thinking of the Kitaev model on the dual lattice compared to Kitaev's original construction, see~\cite{Buerschaper09,SimonBook}.
In this model, the category is based on a group $G$ and the fusion rules simply follow the group multiplication rules. In cases where $G$ is a noncommutative group we have a string-net model with noncommutative fusion rules.   The Kitaev quantum double model is thus a string-net model based on the trivial categorification of $G$, which is known as the category ${\cal C} = $ Vec($G$).   This is an interesting case because there is a Morita equivalence to the commutative fusion category based on the representations of the group ${\cal C'} =$ Rep($G$) where the fusion rules follow the multiplication rules of the group irreducible representations.  Although Vec($G$) and Rep($G$) have the same Drinfeld center, and hence string-net models built from the two are described by the same TQFT, we will see that their nontopological degeneracies are quite different. 

In addition to UFCs based on noncommutative groups, there are other noncommutative fusion categories, such as the Haagerup category~\cite{Asaeda99,Hong08}, which we will discuss in detail below. Construction of string-net models from these categories (and more generally from categories not possessing the tetrahedral symmetry, see e.g.~\cite{SimonBook}) requires some care and is discussed in details, for example, in Refs.~\cite{Lin14,Lan14,Hahn20,Lin21}. This is known as ``generalized string-net models'' and should not be confused with the ``extended string-net models'' of Ref.~\cite{Hu18}.

Our goal is to compute the spectral degeneracies of the generalized string-net models, hence extending the recent results obtained in Ref.~\cite{Vidal22} to any UFC.  Let us stress that the calculation of degeneracies for some specific UFCs has been the subject of several works in the last decade~\cite{Schulz13,Schulz14,Schulz15,Hu12,Hu14}.
Our analysis will begin by constructing the Drinfeld center $\cal Z({\cal C})$  for a string-net model built from any input UFC $\cal C$.   Building the Drinfeld center can in general be quite complicated.  Our approach will rely on the so-called {\it Ocneanu tube algebra}~\cite{Ocneanu94,Ocneanu01}, which gives us all the objects in $\cal Z({\cal C})$ together with their internal degeneracies. Then, using the Moore-Seiberg-Banks formula~\cite{Moore89} supplemented by degeneracy factors we will calculate the degeneracies of each energy levels. This final step requires several nontrivial manipulations which we will emphasize. 
At the end, we will obtain expressions for the spectral degeneracies of generalized string-net models (restricted to the fluxon sector). These will be used in an analysis of the thermal properties of these models in a forthcoming paper. 

Another motivation for the present study is conceptual: it is to clearly understand how much of the physics of the string-net model is captured by the Drinfeld center of the input category and how much is left out. The former may be called topological and the latter nontopological. Here, we will show that beyond the Drinfeld center, one needs quantities called $n_{A,1}$, (see below) extracted from the tube algebra, in order to compute the excited-level degeneracies.

As a final motivation, there are now proposals to simulate simple string-net models on quantum computers (see for example, \cite{Iqbal23} and \cite{Jovanovic23}, which experimental groups are now trying to implement). It therefore seems important to consider such models in detail and to draw out the physical differences that arise from the different ways the Drinfeld centers may be realized.

The paper is structured as follows. 
In Sec.~\ref{sec:fc}, we provide a few basic notions on fusion categories. In Sec.~\ref{sec:gsnm}, we give a short introduction to generalized string-net models, with a particular emphasis on the construction and description of the emergent topological phase.  Section~\ref{sec:spdeg} contains our main result, namely the formula for the spectrum degeneracies in the most general case. We make the link  with previously obtained results in some particular cases. Special cases such as modular or commutative categories are discussed in Sec.~\ref{sec:specialcases}. We further illustrate our results with examples in Sec.~\ref{sec:examples}. In particular, we study the case of the Morita-equivalent categories Rep($S_3$) and Vec($S_3$), corresponding to the same Drinfeld center $\mathcal{Z}(S_3)$ but having different energy-level degeneracies. We also study more involved cases such as the Haagerup, the Tambara-Yamagami and the Haage-Hong categories. Eventually, we conclude and discuss perspectives in Sec.~\ref{sec:conclusion}. In appendices, we give details on the tube algebra (see Appendices~\ref{app:tube} and \ref{app:halfbraiding}), on fluxon identities (see Appendix~\ref{app:condensation}), on a generalized Hamiltonian (see Appendix~\ref{app:Hgen}) and on the Hilbert-space dimension (see Appendix~\ref{app:hilbspacedim}).

%
%
%%%%%%%%%%%%%%%%%%%
%%%%%%%%%%%%%%%%%%%
\section{Background information on fusion categories}
\label{sec:fc}
%%%%%%%%%%%%%%%%%%%
%%%%%%%%%%%%%%%%%%%
%
%

Before starting our discussion, we will need to introduce a few concepts from the theory of TQFTs and fusion categories. Those familiar with the field can skip this section.  A UFC $\mathcal{C}$ is essentially defined by a set of simple objects, a set of fusion rules, and a set of $F$-matrices.  We will now describe each of these in turn. 

The first property of a UFC is the set of simple objects (also called particles, labels, quantum numbers, strings, superselection sectors, ...).  The identity $1$ (also called the vacuum or trivial object) is one of these objects.  Below, we label the (non-identity) simple objects of the UFC with Roman letters $a\in \mathcal{C}$.  

The second property of the UFC is the set of fusion rules. This defines how objects ``multiply" together, such as $a \times b = 2c + f + h$.  These fusion rules are encoded in coefficients $N_{ab}^c$.   In particular, we write
    \beq 
a\times b=\sum_c N_{ab}^c \:c,
\label{eq:fusion}
\eeq 
where the elements $N_{ab}^c$ are just the nonnegative integer coefficients of the object $c$ on the right-hand side of the equation. With only these two properties (simple objects and fusion rules), one has a fusion ring. For a list of small multiplicity-free fusion rings, see Ref.~\cite{AnyonWiki}.
%
%
%%%%%%%%%%%%%
\begin{figure}[t]
\scalebox{.7}{\includegraphics{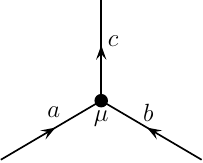}}
\caption{Oriented fusion vertex corresponding to Eq.~(\ref{eq:fusion}). A fusion is ``allowed" if $N_{ab}^c > 0$ where $a$, $b$ and $c$ are in counterclockwise order.  In this case the index at the vertex takes values $\mu = 1, \ldots, N_{ab}^c = N_{ab\bar c}$.  In cases where $N_{ab}^c =1$ we do not write $\mu$.}
\label{fig:orientedvertex}
\end{figure}
%%%%%%%%%%%%%
%
%
If there are $N_\mathcal{C}$ simple objects in the category, we can think $N_{ab}^c$ as being a set of $N_\mathcal{C}$ square matrices $N_a$ which are each $N_\mathcal{C}$-dimensional, with rows and columns indexed by $b,c$.     The quantum dimension $d_a$ is defined to be the largest eigenvalue of $N_a$, which is guaranteed to be positive.  The so-called total quantum dimension is 
\beq
\mathcal{D} = \mbox{$\sqrt{\sum\limits_{a} d_a^2}$}.    
\eeq

The fusion rules are always assumed to be associative
\beq
 a \times (b \times c) = (a \times b) \times c,  \: \text{for all } (a,b,c) \in \cal C.
\eeq

If $a\times b = b\times a, \: \text{for all } (a,b) \in \cal C$,   or equivalently $N_{ab}^c = N_{ba}^c$, we say that the fusion rules are commutative.    Conversely if there exists $(a,b) \in \cal C$ such that $a \times b \neq b \times a$, then we say that the fusion rules are noncommutative. Most studies of string-net models have focused on the simple case of commutative fusion rules with $N_{ab}^c \leqslant 1$ (no fusion multiplicities), but we will be more general. 

Fusion with the identity object $1$ is trivial, meaning that $1 \times a = a \times 1 = a, \: \text{for all } a \in {\cal C}$.    For each simple object $a$ there is a unique dual simple object $\bar a$ such that $N_{a\bar a}^1 = N_{\bar a a}^1 = 1$.   It is possible that $a=\bar a$.    It is always the case that $N_{ab}^c = N_{\bar b \bar a}^{\bar c}$.

It is also sometimes useful to define 
\beq
N_{abc} = N_{ab}^{\bar c}.
\eeq
This notation is convenient because $N_{abc} = N_{bca} = N_{cab}$ even for noncommutative fusion.   

The fusion of objects in the category is represented by diagrams with labeled and directed edges.   For example, fusion of $a$ and $b$ to $c$ is represented by the diagram shown in Fig.~\ref{fig:orientedvertex}.   Note that we have adopted the orientation convention given in Refs.~\cite{Lin21,Hahn20} (one reads $a,b,c$ counterclockwise for the diagram of the vertex $N_{ab}^c$).      Reversing the arrow on a labeled edge changes the edge label to its dual.

\begin{figure}[t]
\scalebox{1.2}{\includegraphics{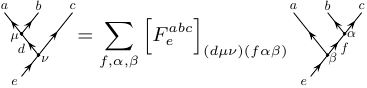}}
 \caption{The space described by the set of diagrams on the left, spanned by different values of $d,\mu,\nu$,  is the same as the space described by the diagrams on the right, spanned by different values of $f,\beta,\alpha$.  The $F$-matrix describes the unitary basis transformation between the two descriptions.}
 \label{fig:Fmove}
\end{figure}

In a UFC, a fusion vertex,  such as the one in Fig.~\ref{fig:orientedvertex}, represents a Hilbert space of dimension $N_{ab}^c$ with basis vectors labeled with greek letter $\mu = 1\ldots N_{ab}^c$. The diagram on the left of Fig.~\ref{fig:Fmove} describes a Hilbert space of dimension $\sum_d N_{d \bar b \bar a} N_{e \bar c \bar d}$ and this space is spanned by the possible values of $d,\mu,\nu$.  On the other hand, the diagram on the right describes a Hilbert space of dimension $\sum_f N_{f \bar c \bar b} N_{e \bar f \bar a}$ and this space is spanned by the possible values of $f,\alpha,\beta$.   Due to the associativity condition on fusion, the dimension of the space in the two descriptions are the same, and the unitary $F$-matrix shown in Fig.~\ref{fig:Fmove} relates the two bases to each other.   There is a consistency condition on the $F$-matrices, known as the pentagon equation~\cite{Kitaev06, Bonderson_thesis, Wang_book, SimonBook}, that assures that multiple changes of basis in complicated diagrams will give a consistent result.   

The essence of the UFC is that it gives us a consistent planar diagram algebra. The values of labeled diagrams can be related to each other via $F$-moves, and any ``closed" diagram (diagrams where all edges end on trivalent vertices at both of their ends) can be evaluated to be a complex number~\cite{SimonBook}. 

\begin{figure}[t]
\includegraphics[width=0.8\columnwidth]{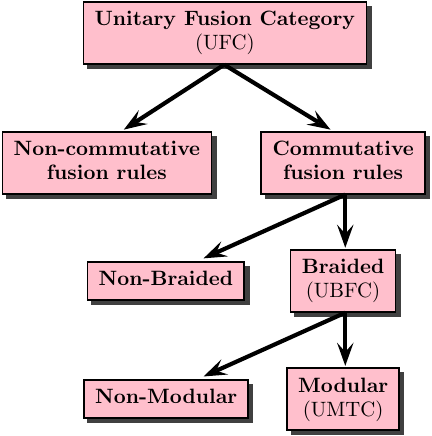}
\caption{A possible classification of unitary fusion categories.} 
\label{fig:schema}
\end{figure}

As mentioned above, the fusion rules of a UFC may be commutative or noncommutative. A simple example of the noncommutative case is when one considers a UFC  built from noncommutative group elements where fusion rules are simply given by the multiplication table of the group considered. A more complicated example of a UFC with noncommutative fusion rules is the Haagerup category $\mathcal{H}_3$~\cite{Asaeda99,Hong08} which we will discuss further below. 
%
%
%%%%%%%%%%%%%%
%%%%%%%%%%%%%%
%
%
When fusion rules are commutative, several types of UFC can further be distinguished as depicted in Fig.~\ref{fig:schema}. 
First, one may ask whether a commutative UFC is braided  or not. A unitary braided fusion category (UBFC) is a UFC obeying the so-called hexagon equations~\cite{Kitaev06, Bonderson_thesis, Wang_book, SimonBook}. These equations stems from consistency condition required when extending the planar diagram rules to diagrams with over- and under-crossings.  Thus (2+1)-dimensional diagrams, corresponding essentially to space-time diagrams for world-lines of anyons, can be given values.   The rules for handling over- and under-crossings (known as $R$-matrix) are described by these hexagon equations.
It may be the case that for a given UFC, there is no consistent solution to these equations, and the UFC is not braided. For instance, the Tambara-Yamagami category TY$_3$~\cite{Tambara98}, which we will discuss further below, is commutative but non braided. 
%
%
%%%%%%%%%%%%%
%%%%%%%%%%%%%
%
%

\begin{figure}[h]
\scalebox{.9}{\includegraphics{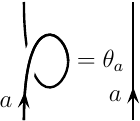}}
\caption{Twist factors $\theta$ appearing in the $T$-matrix.}
\label{fig:twistfactor}
\end{figure}

For UBFCs, important quantities to consider are the so-called $T$-matrix and $S$-matrix. The $T$-matrix, \mbox{$T_{a,b} = \delta_{a,b} \theta_a$}, is simply  a diagonal matrix of the twist factors $\theta_a=\mathrm{e}^{2 \mathrm{i} \pi s_a}$, as shown in Fig.~\ref{fig:twistfactor}, where $s_a$ is the topological spin. 
%
%
%%%%%%%%%%%%%
%%%%%%%%%%%%%
%
%

\begin{figure}[t]
\includegraphics{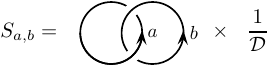}
\caption{The $S$-matrix.}
\label{fig:Smatrix}
\end{figure}

The $S$-matrix is a symmetric matrix of elements \mbox{$S_{a,b}=S_{b,a}$} which evaluate a loop labeled $a$ linked with a loop labeled $b$ and then normalized with the total quantum dimension, as shown in Fig.~\ref{fig:Smatrix}. It gives the exchange statistics of the anyons $a$ and $b$. In the particular case where one of the anyons is the vacuum, the $S$-matrix element takes a simple value in terms of the quantum dimensions:
%
%
%%%%%%%%%%%%%
\beq \label{eq:Squantumdim}
\includegraphics{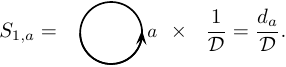}
\eeq
%%%%%%%%%%%%%
%
%
A word of caution is in order. Diagrams in Figs.~\ref{fig:twistfactor} and \ref{fig:Smatrix} as well as in Eq.~(\ref{eq:Squantumdim}) are \textit{evaluated} meaning that they give complex numbers corresponding to quantum amplitudes. Most other diagrams that will appear later in this paper, e.g.,  in Figs.~\ref{fig:twohandled} or \ref{fig:genusg} and in Appendix~\ref{app:hilbspacedim},  are used differently and as a way to compute the dimension of Hilbert spaces, similarly to what is done in Figs. ~\ref{fig:orientedvertex} and ~\ref{fig:Fmove}. The two ways of considering a diagram, i.e.,  either as a quantum amplitude or as a way of counting possible labelings, should be clearly distinguished. For example, in the second case, loops or bubbles can never be contracted, whereas in the first case, an isolated loop of $a$ is evaluated as the quantum dimension $d_a$ [see Eq.~(\ref{eq:Squantumdim})].

If for all $a$ and $b$ there is a single $c$ such that $N_{ab}^c>0$, then the category is said to be Abelian (in the sense that, for an Abelian UBFC, the corresponding objects are all Abelian, i.e., exchange statistics is just a phase) and non-Abelian otherwise. Note that for a fusion category, and in contrast to a group, the notions of commutativity and of Abelianity are distinct. For instance, the Fibonacci UFC is commutative but non-Abelian.

UBFCs can still be split into two families according to whether their $S$-matrix  is unitary, in which case we say the theory is {\it modular}, or not.   A UBFC is modular if it has no transparent particles besides the identity --- where ``transparent" means that it accumulates no phase when braided all the way around any other particle. If a UBFC is modular, it is known as a unitary modular tensor category, or UMTC. A crucial example of a UMTC is the Drinfeld center $\cal{Z}({\cal C})$, which is a theory constructed from any UFC $\cal C$.  We will discuss this example in detail below. Collectively, $T$ and $S$ are then known as the modular matrices.

The $S$-matrix of a UMTC has the crucial property that it simultaneously diagonalizes all of the fusion matrices.  The famed Verlinde formula~\cite{Pasquier87,Verlinde88} states that
%
%
%%%%%%%%%%%%%%%%
\beq
N_{a}= S \Lambda_{a} S^\dagger, \text{ where } (\Lambda_{{a}})_{bc}=\delta_{b,c} \, \frac{S_{a,b}}{S_{1,b}}.
\label{eq:Verlinde}
\eeq
%%%%%%%%%%%%%%%%
%
%
In fact, a similar formula holds true for any commutative UFC (even if non modular or non braided).  One can always define a matrix $\tilde{s}$, sometimes known as the {\it mock} $S$-matrix, that simultaneously diagonalizes all of the fusion matrices as in Eq.~(\ref{eq:Verlinde}).  

A quantity known as the topological central charge $c$ may be defined via
\beq
 \mathrm{e}^{2 \mathrm{i} \pi c/8} = \frac{1}{\mathcal{D}} \sum_a d_a^2 \, \theta_a. 
 \label{eq:centralcharge}
\eeq
%{\red Ju: I guess that $c_a$ should be replaced by $d_a$ in this equation since it is the quantum dimension ? Fixed.  It was a typo}
We say a theory is achiral if $c=0$, even if the theory breaks time-reversal invariance (we will see an example of this below in Sec.~\ref{sec:ty3}). Strictly speaking, Eq.~(\ref{eq:centralcharge}) only defines the topological central charge modulo 8 and one should distinguish it from the chiral central charge (defined without the modulo 8), which determines the thermal Hall conductance~\cite{Kane97,Kitaev06}. However, this distinction will not be important for us. 
%
%
%%%%%%%%%%%%%%%%
%%%%%%%%%%%%%%%%
%
%

%
%
%%%%%%%%%%%%%%%%%%%
%%%%%%%%%%%%%%%%%%%
\section{Generalized string-net models}
\label{sec:gsnm}
%%%%%%%%%%%%%%%%%%%
%%%%%%%%%%%%%%%%%%%
%
%

Inspired by the original construction proposed by Levin and Wen~\cite{Levin05}, but without assuming tetrahedral symmetry, the generalized string-net models~\cite{Lan14,Lin14,Lin21,Hahn20} allow us to generate all topological phases described by the Drinfeld center $\mathcal{Z(C)}$ of a UFC $\mathcal{C}$.  To compute the spectral degeneracies of generalized string-net models, we do not have to explain the detailed construction of the model which can be found in Refs.~\cite{Lin21,Hahn20}.   We will nonetheless give a rough picture of it here. 

The generalized string-net model can be defined on any trivalent graph (see Fig.~\ref{fig:trivalentlattice}) and its construction only requires a (input) UFC $\mathcal{C}$. The Hilbert space $\mathcal{H}$ is defined by the set of all possible allowed edge and vertex labelings.   The trivalent graph is intended to be dual to a triangulation of a genus $g$ two-dimensional orientable manifold.

%%%%%%%%%%%%
%%%%%%%%%%%%
\subsection{The Hamiltonian}
\label{sub:Hamiltonian}
%%%%%%%%%%%%
%%%%%%%%%%%%

The Hamiltonian for a string-net model is written in terms of a set of operators $A_v$, defined at the position of each vertex $v$, and operators $B_p$ defined at the position of each plaquette $p$.   These operators are projectors, meaning that $A_v^2 = A_v$ and $B_p^2 = B_p$ or equivalently that their eigenvalues are 0 and 1 only.  Further these operators all commute with each other: \mbox{$[A_v,A_{v'}] = [A_v,B_p] = [B_p,B_{p'}]=0$}, for all $v,v',p,p'$.

The exactly solvable string-net Hamiltonian is 
\beq
     H = -J_v \!\!\! \sum_{{\rm vertices} \, v} A_v \,\, - J_p \!\!\!\sum_{{\rm plaquettes} \, p} B_p,
     \label{eq:LWH}
\eeq
with $(J_v,J_p) >0$. The $A_v$ operators yield unity if the fusion rules of the input category $\mathcal{C}$ are obeyed at vertex $v$, and give zero otherwise. In a lattice gauge theory analogy, $A_v=1$ corresponds to satisfying a Gauss law at vertex $v$, while $A_v=0$ corresponds to having a ``charge'' at this vertex, i.e., a vertex excitation. The $B_p$ operators can be thought of as assuring that no ``flux" penetrates the plaquette $p$ (the notion of flux being dual to that of charge). The eigenvalue $B_p=1$ corresponds to having no flux in the plaquette $p$ and $B_p=0$ to having a flux, i.e., a plaquette excitation. The detailed form of the $B_p$ operator is a bit complicated, and is given in terms of $F$-symbols of the input UFC. Since we do not need these for our purpose, we refer the interested reader to Refs.~\cite{Hahn20,Lin21} for more details (see Refs.~\cite{Levin05,SimonBook} for an introduction to the simple case with only commutative fusion rules). 

The ground-state space of the model is spanned by all states $|\psi_0 \rangle$ such that $A_v|\psi_0 \rangle=B_p|\psi_0 \rangle=|\psi_0\rangle$.  Excited states $|\psi\rangle$ violate these constraints, i.e.,  either $A_v|\psi \rangle=0$ and/or $B_p|\psi \rangle=0$ for some $(v,p)$.  

\subsection{The output category}
\label{subsec:outcat}
%%%%%%%%%%%%
%%%%%%%%%%%%

The excitations of the system correspond to nontrivial objects of the output category $\mathcal{Z(C)}$, known as the Drinfeld center of $\cal C$ \cite{Lan14}.  Here, we will discuss how to infer their properties from the input category $\mathcal{C}$.   

By construction, the Drinfeld center $\mathcal{Z(C)}$ of a UFC  $\mathcal{C}$ is a UMTC whose objects are built from $\mathcal{C}$. As a UMTC, the output category $\mathcal{Z(C)}$ describes an achiral $(c \mod 8=0)$ anyon theory (see, e.g., Refs.~\cite{Kitaev06,Bonderson_thesis,Wang_book,SimonBook} for more informations).  As mentioned in Ref.~\cite{Lin21}, topological order associated with Drinfeld centers are believed to be the most general class of bosonic topological order compatible with gapped boundaries~\cite{Kitaev12, Lin14,Kong14}.  One way to build the Drinfeld center of a UFC consists in deriving the Ocneanu tube algebra~\cite{Ocneanu94,Ocneanu01} and finding its irreducible representations. The main lines of this construction are summarized in Appendix~\ref{app:tube} and some examples are given in Appendix~\ref{app:halfbraiding}. In the following, we will denote by Roman letters  simple objects of $\mathcal{C}$ and by capital Roman letters simple objects of $\mathcal{Z(C)}$, except for the vacua which are denoted $1$ and ${\bf 1}$, respectively.

Any simple object $A$ of the Drinfeld center is given by
%
%
%%%%%%%%%%%%%%%%%%
\beq
A = \big(\mathop{\oplus}_s  n_{A, s} \, s, \Omega_{A} \big),
\eeq
%%%%%%%%%%%%%%%%%%
%
%
where  $n_{A, s}$ is a nonnegative integer (internal multiplicity) counting the number of times the simple object $s\in \mathcal{C}$ appears in the simple object $A \in \mathcal{Z}(\mathcal{C})$, and where the half-braiding tensor $\Omega_{A}$ gives all braiding properties of $A$ (see Appendices~\ref{app:tube} and \ref{app:halfbraiding} for more details).  Practically, one may see $A$ as a rope made of different types of strands (with a weight $n_{A,s}$ for the strand $s$) endowed with braiding properties defined by $\Omega_{A}$. If the input category $\mathcal{C}$ is commutative, one has $n_{A, s} \in \{0,1\}$. By contrast, for noncommutative $\mathcal{C}$,  $n_{A,s}$ can take integer values larger than 1.
Actually, the $\Omega_A$'s entirely define $\mathcal{Z}(\mathcal{C})$ but they also contain more information since they keep track of the original UFC  ($n_{A,s}$'s can be expressed in terms of these fundamental quantities). Let us also stress that the internal multiplicities $n_{A,s}$'s have nothing to do with the fusion multiplicities discussed in Sec.~\ref{sec:fc}.

Finally, let us mention that an alternative route to the description of the excitations consists in finding operators associated to closed strings which commute with the Hamiltonian (\ref{eq:ham}). Such an approach based on string operators is detailed in Refs.~\cite{Levin05, Lin21} and the relation between the two approaches is discussed in~\cite{Lan14}.

\subsection{Fluxon model}
\label{sub:fluxon}

In this work, we will focus on a somewhat simpler Hamiltonian than the full string-net Hamiltonian of Eq.~(\ref{eq:LWH}), since we shall strictly enforce the vertex constraint. 
This is essentially equivalent to taking $J_v \rightarrow \infty$ in Eq.~(\ref{eq:LWH}) and considering only low-energy physics.  Even more simply we can think of this constraint as restricting the Hilbert space to include only the states that are +1 eigenstates of $A_v$ for all vertices.   Within this restricted Hilbert space and setting $J_p=1$, the Hamiltonian takes the simpler form 
%
%
%%%%%%%%%%%%%
 \beq 
 H = - \sum_{{\rm plaquettes\,} p} B_p .
 \label{eq:ham}
 \eeq 
%%%%%%%%%%%%%
%
%

The ground-state space is entirely contained in this restricted Hilbert space. All excitations of the model that remain within this space are plaquette excitations, and known as {\it fluxons}. These are the excitations we will focus on in this paper. 
In Eq.~\eqref{eq:ham}, $B_p$ is the projector onto the vacuum of $\mathcal{Z(C)}$ in the plaquette~$p$. Consequently, the ground-state energy of the system is given by \mbox{$E_0=-N_{\rm p}$} where $N_{\rm p}$ is the total number of plaquette and a state with $q$ fluxons has an energy \mbox{$E_q=-N_{\rm p}+q=E_0+q$}. 

The Hamiltonian (\ref{eq:ham}) assigns a constant energy penalty of one unit to each plaquette which is not in the vacuum state.  It is easy to generalize Eq.~(\ref{eq:ham}) to a form that instead assigns a different energy to each different type of fluxon excitation.  Similarly, we can also study the case where the energy penalties depend on which particular plaquette of the lattice we are considering.   We briefly address these more general cases in Appendix~\ref{app:Hgen}.

The Hamiltonian (\ref{eq:ham}) is obviously fine tuned compared to more generic cases such as that considered in Appendix~\ref{app:Hgen}. Nonetheless, the techniques we introduce here can be used more widely and our main point is to establish and test our method.

%
%
%%%%%%%%%%%%%%%%%%%
\subsection{Identifying fluxons} 
\label{sub:fluxons}
%%%%%%%%%%%%%%%%%%%
%
%

We will need to determine which objects from $\cal{Z}(\cal{C})$ are actually fluxons.   Using the tube algebra construction, it is straightforward to see that $A$ is a fluxon iff \mbox{$n_{A,1}\neq 0$} (see Appendix~\ref{app:tube}). This definition also includes the trivial fluxon $A=\mathbf{1}$ (vacuum) although, strictly speaking, it is not an excitation. Other fluxons will be called nontrivial in the following. The set of all fluxons in $\cal{Z}(\cal{C})$ will be denoted $\mathcal{F}$, while that of nontrivial fluxons will be denoted $\mathcal{F}^*$. As shown in Appendix \ref{app:condensation}, the vector ${\bf n}_1$, the components of which are $n_{A,1}$, obeys
%
%
%%%%%%%%%%%%%
\begin{eqnarray}
S {\bf n}_1 &=& {\bf n}_1,  \label{eq:identityS} \\ 
 T {\bf n}_1&=&{\bf n}_1,     \label{eq:identityT}
\end{eqnarray}
%%%%%%%%%%%%%
%
%
where $S$ and $T$ are the modular matrices of $\mathcal{Z(C)}$.  In the following, we will make an extensive use of these important relations. Similar equations have appeared in other contexts such as anyon condensation~\cite{Neupert16} or gapped boundaries~\cite{Levin13,Lan15}. Here they are obtained as properties of fluxons (i.e. plaquette excitations). In addition, if the input category $\mathcal{C}$ is commutative (noncommutative), the number $N_\mathcal{F}$ of fluxons equals (is strictly smaller than) the number $N_\mathcal{C}$ of simple objects in $\mathcal{C}$.   Other simple objects (non fluxons) in $\mathcal{Z(C)}$ are associated to violations of the vertex constraint (see Sec.~\ref{sub:Hamiltonian}). At first glance, it may seem strange to worry about these latter excitations since they do not belong to the Hilbert space $\mathcal{H}$. However, we will see that they do play a role in the degeneracies of the energy spectrum. 

Finally, let us make a short comment on Morita-equivalent categories. As mentioned in the introduction, two different input categories may have the same Drinfeld center. Thus, one could naively think that two string-net models built from these two categories have the same plaquette excitations but, in general, this is not true.   Indeed, as discussed above,  multiplicities $n_{A, s}$ do not depend only on $\mathcal{Z}(\mathcal{C})$, but also on $\mathcal{C}$. We will illustrate such a situation in Sec.~\ref{sec:vecrep}.

%
%
%%%%%%%%%%%%%%%%%%%
%%%%%%%%%%%%%%%%%%%
\section{Spectral degeneracies}
 \label{sec:spdeg}
%%%%%%%%%%%%%%%%%%%
%%%%%%%%%%%%%%%%%%%
%
%

The goal of the present paper is to explain how to compute the degeneracy of the $q$-fluxon energy level in a string-net model. As can be anticipated, this degeneracy depends on the surface topology and on the input UFC which, as discussed above, determines the nature of the excitations. Before proceeding further, let us stress that this degeneracy has recently been computed in Ref.~\cite{Vidal22} for any modular input UFC. Here, we aim at giving a general expression valid for any input UFC (see Fig.~\ref{fig:schema} for an overview). 

\subsection{Moore-Seiberg-Banks formula}
\label{subsec:msb}

\begin{figure}[t]
\scalebox{1}{\includegraphics{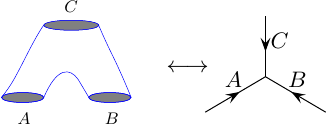}}
\caption{A pants diagram.  Any two-dimensional orientable manifold with $g > 1$ can be decomposed into pants diagrams that are then sewed together. This can be mapped to sewing together trivalent vertices to build a planar fusion diagram. Creating a sphere or a torus in a similar way would require addition of hemispheres. However, computing the degeneracies is equivalent to assigning the identity or vacuum to one (torus) or all (sphere) holes of a pant.}
 \label{fig:pants}
\end{figure}

\begin{figure}[b]
\scalebox{0.85}{\includegraphics{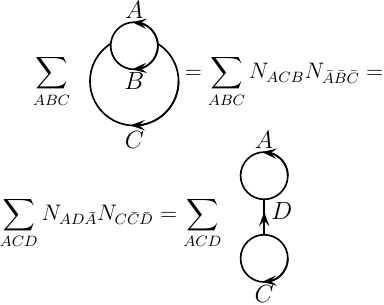}}
    \caption{The ground-state  degeneracy of a two-handled torus. Associativity allows restructuring of the shape of the diagram.  The restructuring of the diagram is basically an $F$-move (see Fig.~\ref{fig:Fmove}) except that here we are only interested in the total dimension of the space described by the diagram. 
    }  
    \label{fig:twohandled}
\end{figure}

We start by considering a TQFT (or UMTC) generically called $\mathcal{U}$ on a 2D orientable manifold of genus $g$.  Here, we have in mind that $\mathcal{U}$ will be a Drinfled center so that we label its simple objects by capital Roman letters.  We can generically decompose this manifold into pants diagrams that are sewn together as in Fig.~\ref{fig:pants}  (assume for now $g>1$ although in the end we do not need this).   Each hole of the pants is labeled with a simple object of $\mathcal{U}$.  The ground-state degeneracy of the resulting manifold is given by all ways consistent with the fusion rules of $\mathcal{U}$ in which these holes can be labeled and then sewn together. This problem of counting the dimension of the ground-state subspace is then mapped to a problem of counting labelings of a planar diagram.   For example, a two-handled torus can be assembled from two pants so that the ground-state degeneracy is given as in Fig.~\ref{fig:twohandled}.   Note that by using associativity of fusion, one can restructure the fusion diagram in multiple equivalent ways, an example of which is shown in Fig.~\ref{fig:twohandled} (see the discussion in Ref.~\cite{Simon13}). 
%
%
%%%%%%%%%%%%%
%%%%%%%%%%%%%
%
%

%
%
%%%%%%%%%%%%%
%%%%%%%%%%%%%
%
%

\begin{figure}
\scalebox{0.95}{\includegraphics{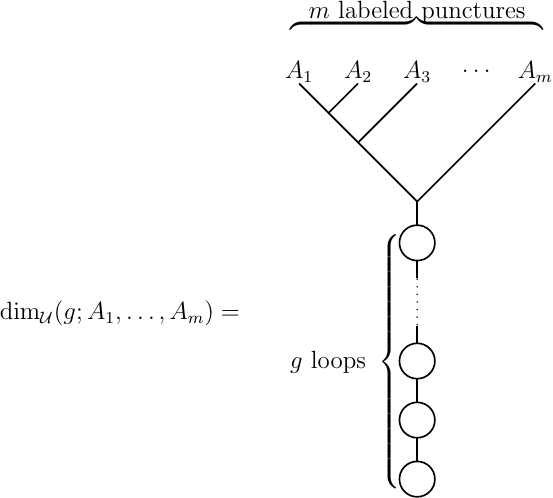}}
    \caption{The fusion diagram for a genus-$g$ surface with $m$ labeled punctures.  Arrows are not drawn on edges for simplicity of presentation. Each vertex in this diagram is a fusion matrix $N$ and each edge is a sum over simple objects in $\mathcal{U}$.  See the example in Fig.~\ref{fig:twohandled}.  For brevity, the sums over unlabeled edges are not written explicitly here.   The sum implied by this diagram can be reduced to Eq.~(\ref{eq:MooreSeiberg}).
       }
 \label{fig:genusg}
\end{figure}

A puncture in the manifold surface may be assigned a simple object type -- essentially any excitation (including the trivial vacuum, which is not really an excitation).  This corresponds to leaving an open, labeled, pant-hole when we sew together our sets of pants.    The most general case we can consider is then a genus $g$ surface with $m$ labeled punctures, and the corresponding fusion diagram is shown in Fig.~\ref{fig:genusg}.  Thus the Hilbert-space dimension $\dim_\mathcal{U}$ associated with this labeled punctured surface just amounts to summing the product of $N$-matrices (one for each trivalent vertex) over all possible internal indices of this diagram.   
%
%
%%%%%%%%%%%%%
%%%%%%%%%%%%%
%
%

While it may look like the sum over all of these internal indices is complicated, using the Verlinde formula and the $S$-matrix of $\mathcal{U}$ [see Eq.~(\ref{eq:Verlinde})], the sum from Fig.~\ref{fig:genusg} can be simplified to the form
%
%
%%%%%%%%%%%%%
\beq
    \mbox{$\dim_{\mathcal{U}}(g; A_1, \ldots, A_m$)} =  \sum_{C} \left[ \prod_{j=1}^m  S_{A_j,C} \right]  S_{\mathbf{1},C}^{2 - 2 g - m} \label{eq:MooreSeiberg},
\eeq
%%%%%%%%%%%%%
%
%
where the sum is performed over all $C \in \mathcal{U}$.  This formula, although it appeared first in the work of Moore and Seiberg~\cite{Moore89}, is  credited to Moore, Seiberg and Banks in a footnote. A special case of this formula was obtained earlier by Verlinde~\cite{Verlinde88}: when there are no punctures in the orientable surface of genus $g$, the Hilbert-space dimension associated with the manifold is just
%
%
%%%%%%%%%%%%%
\beq
\dim_{\mathcal{U}}(g) = \sum_C  S_{\mathbf{1},C}^{2 - 2g}.
 \label{eq:Verlinde2}
\eeq
%%%%%%%%%%%%%
%
%
Another special limit of Eq.~(\ref{eq:MooreSeiberg}) is the case where there are three punctures in a $g=0$ spherical surface (see again Fig.~\ref{fig:pants}).   In this case, one recovers precisely the Verlinde formula (\ref{eq:Verlinde}).
Note that Eq.~(\ref{eq:MooreSeiberg}) has a useful property that if one of the punctures $j$ is labeled by the vaccum, i.e., if $A_j=\mathbf{1}$, it can simply be dropped from the list.   So, for example, one has
%
%
%%%%%%%%%%%%%
\beq
\dim_{\mathcal{U}}(g; A_1,  \dots, A_{m-1}, \mathbf{1}) = \dim_{\mathcal{U}}(g; A_1, \dots, A_{m-1}).  \label{eq:dimdrop}
\eeq
%%%%%%%%%%%%%
%
%
This makes sense since the vacuum quantum number is equivalent to the absence of an excitation.

\subsection{Application to string-net fluxon model}

We now apply the Moore-Seiberg-Banks formula to the string-net model where we only allow fluxon excitations, i.e., we assume all vertex conditions are always satisfied.   
Recall that the string-net model is built from a category $\cal C$ and the emergent TQFT is the Drinfeld center ${\cal Z}({\cal C}) $.  As discussed in Sec.~\ref{sub:fluxon}, fluxons are certain simple objects of ${\cal Z}({\cal C})$. 

Let us consider a string-net model on a graph with $N_\text{p}$ plaquettes embeded on a surface of genus $g$ and a situation in which we have excited $q$ plaquettes by setting  $A_1$, $A_2$... $A_q$ (nontrivial) fluxons labels in these plaquettes. There will still be a degeneracy which comes from the different possibilities how these fluxons can fuse together. 

From the previous Sec.~\ref{subsec:msb}, one might guess that the corresponding degeneracy 
is given by  $\dim_{{\cal Z}({\cal C})}(g;A_1, \ldots, A_{q})$  defined in Eq.~(\ref{eq:MooreSeiberg}), where the UMTC in question here is ${\cal Z}({\cal C})$.  In the case where $\cal C$ has commutative fusion rules, this guess is in fact correct.   However, more generally, the number of states built from the category $\cal C$ on a surface of genus $g$ with fluxons $A_i$ through plaquettes $i =  1 \ldots q$  is given by
%
%
%%%%%%%%%%
\beq
\widetilde{\dim}_{\mathcal{C}} (g;A_1, \ldots, A_{q}) = \dim_{{\cal Z}({\cal C})}(g;A_1, \ldots, A_{q}) \prod_{i=1}^{q} n_{A_{i},1} ,
 \label{eq:degeq}
\eeq
%%%%%%%%%%
 %
 %
where the first term on the right-hand side is given by Eq.~(\ref{eq:MooreSeiberg}) and where we have used the property of Eq.~(\ref{eq:dimdrop}) that plaquettes in the vacuum state do not contribute to the degeneracy. The second term is made from the $n_{A_i,1}$ degeneracy factors assigned to fluxons in Sec.~\ref{sub:fluxons} above.  These internal multiplicities depend on the fluxons  and, as explained in Sec.~\ref{subsec:outcat},  they may be larger than 1 for noncommutative $\cal C$. Equation~(\ref{eq:degeq}) is one of the key statements made in this work.  

The intuition behind Eq.~(\ref{eq:degeq}) comes from an understanding that the string operators for quasiparticles are actually built from the tube algebra (see Appendix \ref{app:tube}).   For each object type in ${\cal Z}({\cal C})$ there are multiple tube types, and this provides the additional degeneracy factor $n_{A, 1}$.  The reason that the second index is 1 is that the end of the string operator should not have a vertex violation in our model (we consider plaquette violations only) which requires the tube end to be labeled with $ 1 \in {\cal C}$ only. An argument for this formula in the language of the tube algebra is given in Appendix \ref{sub:FluxDegLW}.  We have also extensively tested this formula numerically (by exact diagonalization on small systems with different Euler-Poincar\'e characteristics) and found it to always be correct (see Sec.~\ref{sec:examples}).  Further, as we show in Appendix \ref{app:hilbspacedim}, this formula when summed over all possible fluxons on all possible plaquettes, correctly gives the total Hilbert space dimension of the model. 

From Eq.~(\ref{eq:degeq}), it is then easy to compute the total degeneracy of the $q$-fluxon level with energy \mbox{$E_q=-N_\text{p} + q$}, since it is enough to sum over all possibilities for the labels $A_1 \ldots A_q$. 
Up to a combinatorial prefactor given by the binomial coefficient 
%
%
%%%%%%%%%%
$\left(\hspace{-1mm}
\begin{array}{c}
N_{\mathrm p}
\\
q
\end{array}
\hspace{-1mm}
\right)$
%%%%%%%%%%
 %
 %
(which accounts for the number of ways to choose $q$ excited plaquettes among $N_\text{p}$ plaquettes), the degeneracy on a genus $g$ surface with $b=0$ holes is then given by 
%
%
%%%%%%%%%%
\begin{eqnarray}   
D_{ {\cal C}}(g,0,q) &=& \sum_{A_1, \ldots, A_q \in \mathcal{F}^*} \widetilde{\dim}_{\mathcal{C}} (g;A_1, \ldots, A_{q}), \nn \\ 
&=& \sum_{A  \in  {\cal Z}({\cal C} )}S_{\mathbf{1},A}^{2-2g- q}  \prod_{j=1}^q \bigg( \sum_{A_j \in \mathcal{F}^*}S_{A,A_j} n_{A_j,1}\bigg), \nn \\
&=&  \quad  \sum_{A  \in  {\cal Z}({\cal C} )}S_{\mathbf{1},A}^{2-2g- q}  (n_{A,1} - S_{\mathbf{1},A})^{q},
\label{eq:gsqclosed}  
 \end{eqnarray}
 %%%%%%%%%%
 %
 %
where we used Eqs.~(\ref{eq:degeq}) and (\ref{eq:MooreSeiberg}) to go from the first to the second line and Eq.~(\ref{eq:identityS}) from the second to the third. By adding the proper local operator to the Hamiltonian, it is possible to identify the fluxon labels present in the different plaquettes. Therefore, this total degeneracy carries some additional nontopological degeneracy with respect to  Eq.~(\ref{eq:degeq}). Thus in the presence of small amounts of ``disorder", the degeneracy calculated in Eq.~(\ref{eq:gsqclosed}) is genericallly lifted. The above equation for the degeneracy of a level with $q$-fluxons and involving the internal multiplicities $n_{A,1}$ is the main result of the present work.

%
%
%%%%%%%%%%
\subsubsection{Manifold with boundaries} 
%%%%%%%%%%
 %
 %
We now extend the previous results to string-net models on surfaces with boundaries. For a given topological order, there are several types of possible gapped boundaries~\cite{Kitaev12,Lan15}. Here, we only consider smooth boundaries corresponding to the condensation of the fluxons. 
 Consider starting with an orientable manifold without boundary of  genus $g$.  Then we can imagine poking $b$ (potentially large) holes in this manifold to obtain a manifold with boundary.   These holes are essentially (potentially large) plaquettes themselves --- the only distinction between a boundary hole and a regular plaquette is that in Eq.~(\ref{eq:ham}) we sum over only plaquettes that are not these boundary holes, i.e., the energy cost of a boundary hole is zero.

To understand this additional degeneracy, we can start by looking at its contribution in the ground state. The ground-state degeneracy of the string-net model is given by summing over all possible fluxons (including the vacuum fluxon) that can be present in each hole.  All other plaquettes are assumed to be in the ground state, i.e, have the vacuum fluxon quantum number $\mathbf{1}$. In other words, the number $q$ of excited fluxons is $0$.  
 As mentioned above, if a plaquette is known to be in the vacuum state, it does not contribute to any degeneracy. The ground-state degeneracy is  
%
%
%%%%%%%%%%
\begin{eqnarray} 
D_{ {\cal C}}(g,b,0)  
&=& \sum_{A_1, \ldots, A_b \in \mathcal{F}} \widetilde{\dim}_{\mathcal{C}} (g;A_1, \ldots, A_{b}),\nn \\
&=& \:\:\: \sum_{A  \in  {\cal Z}({\cal C} )}S_{\mathbf{1},A}^{2 - 2g - b}  n_{A,1}^{b}.
 \label{eq:gsdbound}  
 \end{eqnarray} 
 Note that, apart from the substitution of $b$ with $q$, the only difference between Eq.~(\ref{eq:gsdbound}) and Eq.~(\ref{eq:gsqclosed}) comes from the fact that we sum either on $\mathcal{F}$ or on $\mathcal{F^*}$. Equation~(\ref{eq:gsdbound})  should be considered as a generalization of Eq.~(\ref{eq:Verlinde2}) to the case with boundaries and with internal multiplicities. In the absence of boundary, i.e., for $b=0$, one recovers Eq.~(\ref{eq:Verlinde2}):
\begin{eqnarray}  
D_{ {\cal C}}(g,0,0) &=&  \dim_{\mathcal{Z}(\mathcal{C})} (g) = \sum_{A  \in  \mathcal{Z}(\mathcal{C})} S_{\mathbf{1},A}^{2-2g},
 \label{eq:gsdbound2}  
 \end{eqnarray}
%%%%%%%%%%
 %
 %
where, by convention, we set $0^0=1$. 

We are now ready to compute the degeneracies of the excited states in the most general situation. This essentially merges the calculations of the previous two sections.   We choose $q$ plaquettes that can have any non-vacuum fluxon (punctures labeled from $1$ to $q$), whereas the $b$ boundaries (punctures labeled from $q+1$ to $b$) can have any fluxon including the vacuum. 

Up to the binomial factor 
%
%
%%%%%%%%%%
$\left(\hspace{-1mm}
\begin{array}{c}
N_{\mathrm p}
\\
q
\end{array}
\hspace{-1mm}
\right)$
%%%%%%%%%%
 %
 %
 discussed before Eq.~(\ref{eq:gsqclosed}), the degeneracy of the excitation level, \mbox{$E_q=-N_\text{p} + q$}, is then
%
%
%%%%%%%%%
\begin{eqnarray}  
\label{eq:result2}
D_{ {\cal C}}(g,b,q) &=& \sum_{ \scriptsize \begin{array}{c}  A_1, \ldots, A_{q} \in \mathcal{F}^* \\ 
 A_{q+1}, \ldots, A_{q+b} \in \mathcal{F} \end{array} }
\!\!\!\!\!\!\!\!\!\!\!\!\! \widetilde{\dim}_{\mathcal{C}} (g;A_1, \ldots, A_{q+b}), \\ 
&=& \sum_{A  \in  {\cal Z}({\cal C} )}S_{\mathbf{1},A}^{2 - 2g - q -b}  (n_{A,1} - S_{\mathbf{1},A})^{q} \,\, n_{A,1}^b, \quad \quad
\label{eq:result3}
 \end{eqnarray}
 %%%%%%%%%
 %
 %
which coincides with Eq.~(\ref{eq:gsqclosed}) in the compact case \mbox{($b=0$)}, as it should.

%%%%%%%%%%
%%%%%%%%%%
\subsubsection{Total Hilbert-space dimension}
%%%%%%%%%%
%%%%%%%%%%

To end this section, let us calculate the total dimension of the Hilbert space $\mathcal{H}$ of a string-net model built from a category $\cal C$ on an orientable surface of genus $g$ with $N_\text{p}$ plaquettes,  $b$ boundaries, and without vertex defects.    This is the total Hilbert-space dimension remaining at finite energy once we have taken  $J_v$ to infinity in the Hamiltonian Eq.~(\ref{eq:LWH}).   One approach to this is to sum over the degeneracy of the string-net model for all possible fluxon states (including the vacuum $\mathbf{1}$) being present in all ($N_\text{p}+b$) ``punctures''.   Thus, we have
%
%
%%%%%%%%%%
\begin{eqnarray}
\dim \mathcal{H} &=&\hspace{-6mm}\sum_{A_1, \ldots, A_{N_\text{p} +b}  \in {\cal Z}({\cal C} )} \hspace{-6mm}\widetilde{\dim}_{\mathcal{C}} (g;A_1, \ldots, A_{N_\text{p} +b}), \quad \\
&=&  \sum_{A \in  {\cal Z}({\cal C} )}S_{\mathbf{1},A}^{2 - 2g - (N_\text{p} +b)}  n_{A,1}^{N_\text{p}+b},\\
&=&D_{ {\cal C}}(g,N_\text{p}+b,0) , 
\end{eqnarray}
%%%%%%%%%%
%
%
 where we have used Eq.~(\ref{eq:degeq}) and the fact that $n_{A,1}$ is nonzero only if $A$ is a fluxon (including the identity), i.e., 
 $A \in \mathcal{F}$. 

For a trivalent graph, one further has $N_\text{e} = \frac{3}{2} N_\text{v}$, where $N_\text{e}$ and $N_\text{v}$  denote  the number of edges and the number of vertices, respectively. The  Euler-Poincar\'e characteristic on a genus-$g$ surface with $b$ boundaries,
is then given by 
%
%
%%%%%%%%%%
\beq
\chi = 2 - 2 g -b = N_\text{p} - N_\text{e} + N_\text{v}=  N_\text{p} - \frac{1}{2} N_\text{v}.
\label{eq:Euler}
\eeq
Thus, we obtain the delightfully simple result for the Hilbert-space dimension of any generalized string-net model:
%
%
%%%%%%%%%%
\beq 
\dim \mathcal{H}=  \sum_{A  \in  {\cal Z}({\cal C} )}S_{\mathbf{1},A}^{-N_\text{v}/2}  n_{A,1}^{N_\text{p}+b}.
 \label{eq:totaldegenresult}
 \eeq
 %%%%%%%%%%
%
%
One straightforwardly sees that for commutative UFCs ($n_{A,1}=0$ or $1$), this dimension only depends on the number of vertices whereas, in the noncommutative case for which $n_{A,1}$ can be larger than 1, $\dim \mathcal{H}$ also depends on $N_\text{p}$. Hence, in this latter case, the Hilbert-space dimension is sensitive to the surface topology. If the input category $\mathcal{C}$ is a UMTC, one recovers the result of~\cite{Hermanns14, Vidal22}:
%
%
%%%%%%%%%%
 \beq 
\dim \mathcal{H}=  \sum_{j  \in  {\cal C}}\left(\frac{\mathcal{D}_\mathcal{C}}{d_j}\right)^{N_\text{v}},
 \eeq
%%%%%%%%%%
%
%
where $\mathcal{D}=\mathcal{D}_\mathcal{C}^2$ is the total quantum dimension of $\mathcal{Z(C)}$ and $\mathcal{D}_\mathcal{C}$ that of $\mathcal{C}$.
Replacing $N_\text{p}+b$ by $2-2g+N_\text{v}/2$ and keeping in mind that $S_{\mathbf{1},A}=d_A/\mathcal{D}$ and $n_{A,1}\leq d_A$, one gets, in the thermodynamic limit:
%
%
%%%%%%%%%%
 \beq 
\dim \mathcal{H} \simeq  \mathcal{D}^{N_\text{v}/2} \sum_{{\rm pure} \: A} n_{A,1}^{2-2g},
 \label{eq:dimHscaling}
 \eeq
%%%%%%%%%%
%
%
 where ``pure" fluxons are fluxons for which $n_{A,1}=d_A$.

Another way of obtaining this result is to return to the original string-net graph (as shown in Fig.~\ref{fig:trivalentlattice}) and label all possible edges with simple objects of $\cal C$ in all possible ways such that all the vertex conditions are satisfied (and counting fusion multiplicities $N_{abc}$ at each satisfied vertex).   The general proof that this counting gives the same result as Eq.~(\ref{eq:totaldegenresult}) is given in Appendix~\ref{app:hilbspacedim} for the case $b = 0$. 

%
%
%%%%%%%%%%%%%%%%%%%
%%%%%%%%%%%%%%%%%%%
\section{Special cases}
\label{sec:specialcases}
%%%%%%%%%%%%%%%%%%%
%%%%%%%%%%%%%%%%%%%
%
%

In the previous sections, all results were given in terms of data from the Drinfeld center and the tube algebra. Yet, for some special cases, we can also write Eq.~(\ref{eq:result3}) completely or partially in terms of data from the input category. We will discuss these cases here, namely when the input category is a UMTC and when the input category is commutative. 

%
%
%%%%%%%%%%%%%%%%%%%
\subsection{Modular categories} 
%%%%%%%%%%%%%%%%%%%
%
%
When the input category of the string-net model is modular, it is possible to compute the degeneracies of the excited levels by using only the fusion properties of the input category. The derivation of this equation is presented in Ref.~\cite{Vidal22}, where the partition function is computed from a microscopic point of view for string-net models built from an input category $\mathcal{C}$ which is a UMTC. We aim here at making the link between the results of the previous sections and the ones of Ref.~\cite{Vidal22}. 
In the case where the input category $\mathcal{C}$ is a UMTC, the Drinfeld center is the product of two copies of $\mathcal{C}$ of opposite chiralities~\cite{Muger03}: 
%
%
%%%%%%%%%%%%%
\beq 
\mathcal{Z}(\mathcal{C})=\mathcal{C}\times\bar{\mathcal{C}}.
\eeq
%%%%%%%%%%%%%
%
%
In this case, any object $A$ of the center can be represented by a couple $(i, j)$ of objects $i \in \mathcal{C}$ and $j \in \bar{\mathcal{C}}$. The $S$-matrix of the Drinfeld center $\mathcal{Z}(\mathcal{C})$ has a simple expression in terms of the $S$-matrix of the input category $\mathcal{C}$ (for clarity, we will denote them respectively $S$ and $s$):
%
%
%%%%%%%%%%%%%
\beq
S_{A, B}=S_{(i, j), (p, q)}=s_{i,p}s^*_{j,q}.
\eeq 
%%%%%%%%%%%%%
%
%
A particle $A =(i,j)$ of $\mathcal{Z}(\mathcal{C})$ is a fluxon iff  $i=j$ and $n_{A,1}=1$, so that $S_{\mathbf{1},A}=s_{1,i}^2$. In all other cases, $n_{A, 1}=0$. Thus, when $b=0$, we can rewrite \eqref{eq:result2} as 
%
%%%%%%%%%%%%%
\begin{eqnarray}
D_{ {\cal C}}(g,0,q) =&& \hspace{-4mm}\sum_{i, j \in \mathcal{C}}(s_{1, i}s_{1, j})^{\chi}\left(\frac{\delta_{i,j}}{s_{1,i}s_{1 j}}-1\right)^q, \nn \\
=&& \hspace{-4mm} 
(-1)^q \Big\{\sum_{i\in \mathcal{C}} s_{1, i}^{2\chi}[(1-s_{1, i}^{-2})^q-1] +D_{ {\cal C}}(g,0,0)\Big\},\nn \\
\label{julien}
\end{eqnarray} 
%%%%%%%%%%%%%
%
%
where $D_{ {\cal C}}(g,0,0) =\bigg(\sum_{i\in\mathcal{C}}s_{1,i}^{\chi}\bigg)^2$ is the ground-state degeneracy. Up to the usual binomial factor, Eq.~(\ref{julien}) is exactly Eq.~(10) in Ref.~\cite{Vidal22}.

%
%
%%%%%%%%%%%%%%%%%%%
\subsection{Commutative categories}
%%%%%%%%%%%%%%%%%%%
%
%

The previous equations do not hold in the case where the input category is commutative but not modular, since we cannot define a modular $S$-matrix. 
However, for any commutative input category, we can always find a unitary matrix $\tilde{s}$, called the mock $S$-matrix, that simultaneously diagonalizes its fusion matrices $N_i$. A Verlinde-like formula (\ref{eq:Verlinde}) holds between the $N_i$ and $\tilde{s}$, see Eq.~(\ref{eq:Verlindemock}). The columns of the matrix $\tilde{s}$ can be indexed by the fluxons $A$ in $\mathcal{Z}(\mathcal{C})$ (for details on the mock $S$-matrix and the ``mixed'' notation for the matrix elements $\tilde{s}_{i,A}$ with $i\in \mathcal{C}$ and $A\in \mathcal{F}$, see Appendix~\ref{sec:mock}). In particular, if $A$ is a fluxon, $S_{\mathbf{1},A}=d_A/\mathcal{D}=\tilde{s}_{1,A}^2$. Also, for any commutative UFC, the multiplicities $n_{A, 1}$ only take values 0 or 1, and the number of fluxons is equal to the number of simple objects in the input category. With these insights at hand, we can rewrite Eq.~(\ref{eq:result3}) as
%
%
%%%%%%%%%%%%%
\beq
D_{ {\cal C}}(g,0,q)=
(-1)^q \Big\{\sum_{A\in \mathcal{F}} \tilde{s}_{1, A}^{2\chi}[(1-\tilde{s}_{1, A}^{-2})^q-1] +D_{ {\cal C}}(g,0,0)\Big\},
\label{finaldegstilde}
\eeq
%%%%%%%%%%%%%
%
%
which is very similar to Eq.~(\ref{julien}) with $s_{1,i}$ replaced by $\tilde{s}_{1,A}$ and fluxons labeled by $A\in\mathcal{F}$ instead of $i \in \mathcal{C}$. However, in the present case, there is no general formula for the ground-state degeneracy  $D_{ {\cal C}}(g,0,0)$ in terms of $\tilde{s}$. One could think of using Eq.~(\ref{eq:gsdbound2}) to compute the ground-state degeneracy with $S_{\mathbf{1},A}=\tilde{s}_{1,A}^2$, but this relation only holds for fluxons and not for all $A\in \mathcal{Z(C)}$.

Generally speaking, it is not possible to compute the ground-state degeneracy $D_{ {\cal C}}(g,0,0)$ simply from the knowledge of the fusion rules of the input category. In fact,  two input categories that have the same fusion rules but different $F$-symbols generally lead to two different Drinfeld centers. In particular, the number of simple objects of the Drinfeld centers need not be the same~\cite{Joost}. And even if the numbers match, the corresponding quantum dimensions $d_A$ need not be equal and therefore the level degeneracies can differ.
However, there still is a way to write $D_{ {\cal C}}(g,0,0)$ only in terms of the input data. This way, which is a bit cumbersome, requires the use of the $F$-symbols of the category (see Ref.~\cite{Hu12}).

%
%
%%%%%%%%%%%%%%%%%%%
%%%%%%%%%%%%%%%%%%%
\section{Examples}
 \label{sec:examples}
%%%%%%%%%%%%%%%%%%%
%%%%%%%%%%%%%%%%%%%
%
%

This section is devoted to nontrivial examples that illustrate the formulas obtained in the previous sections. Before discussing some specific examples, let us give some general results that are valid for any input category ${\cal C}$. In Table~\ref{tab:gen}, we give the degeneracies $D_{ {\cal C}}(g,b,q)$ [see Eq.~(\ref{eq:result3})] of the first three energy levels  for various simple surface topologies.
%
%
%%%%%%%%%%%%%
\begin{table}[h!]
\center
\begin{tabular}{ | c || c | c | c | c | } 
\hline
 $D_{{\mathcal C}}$ & $(g,b)=(0,0)$ & $(0,1)$ & $(0,2)$ & $(1,0)$   \\ \hline
 \hline
 $q=0$  &  1  & 1 & $N_\mathcal{C}$ & $N_\mathcal{Z}$  \\ \hline   
1 & 0 & $N_\mathcal{C}-1$ &  &    \\ \hline 
2 & $N_\mathcal{C}-1$ & &  &   \\ \hline
\end{tabular}
\caption{Degeneracies $D_{ {\cal C}}(g,b,q)$ of the $q$-th excited level (row index: $q=0,1,2$) for various surface topologies characterized by their genus $g$ and their $b$ boundaries [column index $(g,b)$]. The number of simple objects  in the input category $\mathcal{C}$ and in its Drinfeld center $\mathcal{Z}(\mathcal{C})$ are denoted by $N_\mathcal{C}$ and $N_\mathcal{Z}$, respectively. Empty entries depend on the input category details.}
\label{tab:gen}
\end{table} 
%%%%%%%%%%%%%
%
%

To obtain these results, we used  Eq.~(\ref{eq:identityS}) which leads to 
%
%
%%%%%%%%%%%%%
\beq
\sum_{A  \in  {\cal Z}({\cal C} )}S_{\mathbf{1},A} n_{A,1}=n_{\mathbf{1},1}=1,
\eeq
%%%%%%%%%%%%%
%
%
 and the following identity:
%
%
%%%%%%%%%%%%%
 \beq
 \sum_A n_{A,1}^2=N_\mathcal{C},
 \eeq
%%%%%%%%%%%%%
%
%
which is shown in Appendix~\ref{app:tube}.

Whenever possible, i.e., if the Hilbert space is not too large, we checked these degeneracies by exact diagonalizations of the Hamiltonian (\ref{eq:ham}) on trivalent graphs with the corresponding topology. For instance, starting from a cube ($g=b=0$, and $N_\text{v}=8$), we can build a ``disk" ($g=0$, $b=1$) by removing one plaquette operator in the Hamiltonian or a ``cylinder" ($g=0$, $b=2$) by removing two opposite plaquette operators.

%
%
%%%%%%%%%%%%%%%%%%%
\subsection{Rep($S_3$) and Vec($S_3$) categories}
\label{sec:vecrep}
%%%%%%%%%%%%%%%%%%%
%
%

We present here two string-net models built from two Morita-equivalent categories: Rep($S_3$) and Vec($S_3$), where $S_3$ is the symmetric group on a set of 3 elements (symmetry group of the equilateral triangle). They correspond respectively to the category of irreducible representations of $S_3$ and the category of the elements of the group $S_3$.

The category $\mathcal{C}$ = Rep($S_3$) contains $N_\mathcal{C}=3$  simple objects $\lbrace 1, 2, 3 \rbrace$ with quantum dimensions $\{1,1,2\}$, which correspond to the irreducible representations of $S_3$.  Fusion rules and $F$-symbols can be found, e.g., in Ref.~\cite{Levin05}. The fusion rules are commutative but the category is non-Abelian, braided, and non modular. 

The category $\mathcal{C}$ = Vec($S_3$) contains $N_\mathcal{C}=6$ simple objects \mbox{$\lbrace e, \zeta, \zeta^2, \tau, \tau\zeta,\tau\zeta^2\rbrace$}, which are the elements of the group $S_3$, ($e$ is the identity element, $\zeta$ is a $2\pi/3$ rotation and $\tau$ is a mirror). The fusion rules are simply the multiplication rules of the group. All elements have quantum dimension \mbox{$d_s=1$}. The fusion rules are noncommutative but the category is Abelian. 
 
The two categories are Morita-equivalent and lead to the same Drinfeld center $\mathcal{Z}(S_3)$. The latter contains $N_\mathcal{Z}=8$ simple objects, denoted by $\lbrace A, B, C, D, E, F, G, H \rbrace$, see Ref.~\cite{Beigi11}. The quantum dimensions are  $\lbrace 1, 1, 2, 3, 3, 2, 2, 2\rbrace$, and the total quantum dimension is $\mathcal{D}=6$. 

While the two categories share the same center $\mathcal{Z}(S_3)$, the tube algebras are different (see Appendix~\ref{app:halfbraiding}). In particular, the internal multiplicities of the quasiparticles, $n_{J, s}$ differ. We give them as rectangular matrices $n_{\mathcal{C}}$ with $N_\mathcal{C}$ columns and $N_\mathcal{Z}$ rows: 
%
%
%%%%%%%%%%%%%
\beq
n_{\text{Rep}(S_3)} = 
\begin{pmatrix} 
1 & 0 & 0\\ 
0 & 1 & 0\\
0 & 0 & 1\\
1 & 0 & 1\\
0 & 1& 1\\
1 & 1 & 0\\
0 & 0 & 1\\
0 & 0 & 1\\
\end{pmatrix},
\eeq 
%%%%%%%%%%%%%
%
%
%
%
%%%%%%%%%%%%%
\beq
n_{\text{Vec}(S_3)}=
\begin{pmatrix} 
1 & 0 & 0 & 0 & 0 & 0\\
1 & 0 & 0 & 0 & 0 & 0\\
2 & 0 & 0 & 0 & 0 & 0\\
0 & 0 & 0 & 1 & 1 & 1 \\
0 & 0 & 0 & 1 & 1 & 1 \\
0 & 1 & 1 & 0 & 0 & 0 \\
0 & 1 & 1 & 0 & 0 & 0 \\
0 & 1 & 1 & 0 & 0 & 0 \\
\end{pmatrix}.
\label{eq:nvec}
\eeq 
%%%%%%%%%%%%%
%
%
Quantum dimensions of the quasiparticles in the center and the multiplicities are the only quantities we need to compute the degeneracies of any energy level with Eq.~(\ref{eq:result3}). 
In both cases there are three types of fluxons (quasiparticles $J$ with $n_{J,1}> 0$) but they correspond to two different subsets of the simple objects of $\mathcal{Z}(S_3)$: $A, D, F$ for Rep($S_3$), and $A, B, C$ for Vec($S_3$) (note that $n_{C,1}=2$). As a consequence, the spectral degeneracies of the two models are different (see Tables~\ref{tab:rep} and \ref{tab:vec}). 
%Some numerical calculations were performed on small systems and in the link basis to confirm the formulas and are given in .
%
%
%%%%%%%%%%%%%
\begin{table}[h!]
\center
\begin{tabular}{ | c || c | c | c | c | } 
\hline
$D_{ {\rm Rep}(S_3)}$ & $(g,b)=(0,0)$ & $(0,1)$ & $(0,2)$ & $(1,0)$   \\ \hline \hline
$q=0$ & 1 & 1 & 3 & 8  \\ \hline   
1 & 0 & 2 & 8 & 3   \\ \hline 
2 & 2 & 6 & 30 & 35   \\ \hline   
3 &4 & 24  & 134 & 129  \\ \hline      
4 & 20 & 110 & 642 & 647  \\ \hline 
\end{tabular}
\caption{Degeneracies of the $q^{th}$ excited state of a string net built from Rep($S_3$) for various surface topologies up to $q=4$.  Here $g$ is the genus and $b$ is the number of boundaries.   }
\label{tab:rep}
\end{table}
%%%%%%%%%%%%%
%
%

%
%
%%%%%%%%%%%%% 
\begin{table}[h!]
\center
\begin{tabular}{ | c || c | c | c | c |} 
\hline
$D_{{\rm Vec}(S_3)}$ & $(g,b)=(0,0)$ & $(0,1)$ & $(0,2)$ & $(1,0)$  \\ \hline\hline
$q=0$ & 1 & 1 & 6 & 8  \\ \hline   
1 & 0 & 5 & 30 & 10 \\ \hline 
2 & 5 & 25 & 150 & 80  \\ \hline   
3 & 20 & 125 & 750 & 370  \\ \hline      
4 & 105 & 625 & 3750 & 1880 \\ \hline 
\end{tabular}
\caption{Degeneracies of the $q^{th}$ excited state of a string net built from Vec($S_3$)  for various surface topologies up to $q=4$.   Here $g$ is the genus and $b$ is the number of boundaries.}
\label{tab:vec}
\end{table} 
%%%%%%%%%%%%%
%
%

The dimension of the Hilbert space [see Eq.~(\ref{eq:totaldegenresult})] is
%
%
%%%%%%%%%%%%%
\beq
\dim \mathcal{H} = \left(1+3^{-N_\text{v}/2}+2^{-N_\text{v}/2}\right) \sqrt{6}^{N_\text{v}},
\eeq 
%%%%%%%%%%%%%
%
%
for Rep($S_3$) and
%
%
%%%%%%%%%%%%%
\beq
\dim \mathcal{H}= (1+1+2^{2-2g}) \sqrt{6}^{N_\text{v}},
\eeq 
%%%%%%%%%%%%%
%
%
for Vec($S_3$). We provide a few values of  $\dim \mathcal{H}$ in Table~\ref{hilbertdim} for some simple trivalent graphs.
%
%
%%%%%%%%%%%%%
\begin{table}[h!]
\center
\begin{tabular}{ | c || c | c | c | c | c | c | } 
\hline
$\dim \mathcal{H}$ & $(g,N_\text{v})=(0,2)$ & (1,2) & (0,4) & (1,4) & (0,6) & (1,6)  \\ \hline \hline
Rep($S_3$) & 11 & 11 & 49 & 49 & 251 & 251 \\ \hline   
Vec($S_3$) & 36 & 18 & 216 & 108 & 1296 & 648 \\ \hline 
TY$_3$ & 18 & 18 & 90 & 90 & 486 & 486 \\ \hline
$\mathcal{H}_3$ & 63 & 45 & 1431 & 1323 & 46494 & 45846  \\ \hline
\end{tabular}
\caption{Hilbert-space dimension $\dim \mathcal{H}$ for some simple compact surfaces ($b=0$) and for some input categories. Each row corresponds to a given category and each column is indexed by $(g,N_\text{v})$ with $g$ the genus and $N_\text{v}$ the number of vertices.}
\label{hilbertdim}
\end{table} 
%%%%%%%%%%%%%
%
%

%We have checked numerically that the twofold level degeneracy ($n_{C,2}=2$) associated to the $C$ fluxon for the Vec($S_3$) string-net model can be lifted by a local perturbation in the Hamiltonian. This confirms that such an internal degeneracy is non topological.
%, unlike that related to $S_{1,\alpha}$.

The fusion matrices of the commutative category Rep$(S_3)$ are simultaneously diagonalized by the following mock $S$-matrix:
\beq
\tilde{s}=\frac{1}{\sqrt{6}}\left(
\begin{array}{ccc}
1&\sqrt{3}&\sqrt{2}\\
1&-\sqrt{3}&\sqrt{2}\\
2 &0&-\sqrt{2}
\end{array}
\right) .
\eeq
This matrix contains the transformation $\tilde{s}_{i,J}$ between the simple objects $i \in \{1,2,3\}=\mathcal{C}$ of the input category and the fluxons $J\in \{A,D,F\}=\mathcal{F}$ of the output category $\mathcal{Z(C)}$. Here, it was chosen such that the first row only contains strictly positive elements, in which case, one has $\tilde{s}_{1,J}=\sqrt{d_J/\mathcal{D}}$ and $\tilde{s}_{i,\mathbf{1}}=d_i/\sqrt{\mathcal{D}}$ (see Appendix~\ref{sec:mock}).

In conclusion, Rep($S_3$) and Vec($S_3$) are two Morita-equivalent UFCs that correspond to the same Drinfeld center but give different Hilbert spaces and spectral degeneracies (and hence have different partition functions).

Note that, in contrast to the case presented here, there also exist UFCs corresponding to different Drinfeld centers that can give rise to the same energy spectrum and level degeneracies. This is the case, e.g., for  Vec($\mathbb{Z}_2$) and the semion theory Vec$^\omega (\mathbb{Z}_2)$ (nontrivial cocycle of $\mathbb{Z}_2$). 

%
%
%%%%%%%%%%%%%%%%%%%
\subsection{Tambara-Yamagami category TY$_3$}
%%%%%%%%%%%%%%%%%%%
%
%
\label{sec:ty3}

The Tambara-Yamagami category for $\mathds{Z}_3$ (notated TY$_3$) is interesting as it breaks the tetrahedral symmetry and therefore can only be realized by the generalized string-net model (see for example~\cite{Lin21}) and not by the original Levin-Wen model~\cite{Levin05}. It is commutative (but non-Abelian and nonbraided), so that $n_{A,1}=1$ for all fluxons. It has \mbox{$N_\mathcal{C}=4$} simple objects $\{ 1, 2, 3, \sigma\}$, with quantum dimensions $\{1,1,1,\sqrt{3}\}$. There are two different solutions to the pentagonal equations indexed by $p=\pm 1$ (see~\cite{Lin21} for more details on the $F$-symbols). The Drinfeld center $\mathcal{Z}({\rm TY}_3)$ contains $N_\mathcal{Z} =15$ elements $\{1,...,15\}$ with dimensions \mbox{$\{1,1,1,1,1,1,2,2,2,\sqrt{3},\sqrt{3},\sqrt{3},\sqrt{3},\sqrt{3},\sqrt{3}\}$} and the total quantum dimension is $\mathcal{D}=6$. For the $p=1$ model, the internal multiplicities are: 
%
%
%%%%%%%%%%%%%
\beq
n_{{\rm TY}_3}=
\begin{pmatrix} 
1 & 0 & 0 & 0\\ 
1 & 0 & 0 & 0\\
0 & 1 & 0 & 0\\
0 & 1 & 0 & 0\\
0 & 0 & 1 & 0\\
0 & 0 & 1 & 0\\
1 & 1& 0 & 0\\
1 & 0 & 1 & 0\\
0 & 1 & 1 & 0\\
0 & 0 & 0 & 1\\
0 & 0 & 0 & 1\\
0 & 0 & 0 & 1\\
0 & 0 & 0 & 1\\
0 & 0 & 0 & 1\\
0 & 0 & 0 & 1\\
\end{pmatrix}.
\eeq 
%%%%%%%%%%%%%
%
%
The fluxons are therefore $\mathbf{1},2,7$ and $8$ and, using Eq.~(\ref{eq:result3}), we can easily compute the degeneracies. Some examples are given in Table~\ref{tab:ty3}.
%
%
%%%%%%%%%%%%%
\begin{table}[h!]
\center
\begin{tabular}{| c || c | c | c | c |} 
\hline
$D_{{\rm TY}_3}$ & $(g,b)=(0,0)$ & $(0,1)$ & $(0,2)$ & $(1,0)$  \\ \hline \hline
$q=0$ & 1 & 1 & 4 & 15  \\ \hline   
1 & 0 & 3 & 14 & 3 \\ \hline 
2 & 3 & 11 & 58 & 69  \\ \hline   
3 & 8 & 47 & 266 & 255  \\ \hline      
4 & 39 & 219 & 1282 & 1293 \\ \hline \end{tabular}
\caption{Degeneracies of the $q^{th}$ excited state of a string net built from  TY$_3$  for various surface topologies up to $q=4$. Here $g$ is the genus and $b$ is the number of boundaries. }
\label{tab:ty3}
\end{table} 
%%%%%%%%%%%%%
%
%

 The Hilbert-space dimension [see Eq.~(\ref{eq:totaldegenresult})]  is
%
%
%%%%%%%%%%%%%
\beq
\dim \mathcal{H}= 2\left(1+2^{-N_\text{v}/2}\right) \sqrt{6}^{N_\text{v}},
\eeq 
%%%%%%%%%%%%%
%
%
and a few examples are given in Table~\ref{hilbertdim}. Interestingly, this topological order breaks time-reversal symmetry~\cite{Lin21} but still has a vanishing topological central charge \mbox{($c \mod 8=0$)}. 

The fusion matrices of TY$_3$ are simultaneously diagonalized by the following mock $S$-matrix:
\beq
\tilde{s}=\frac{1}{\sqrt{6}}\left(
\begin{array}{cccc}
1&1&\sqrt{2}&\sqrt{2}\\
1&1&\sqrt{2}\mathrm{e}^{\mathrm{i} 2\pi/3}&\sqrt{2}\mathrm{e}^{-\mathrm{i} 2\pi/3}\\
1&1&\sqrt{2}\mathrm{e}^{-\mathrm{i} 2\pi/3}&\sqrt{2}\mathrm{e}^{\mathrm{i} 2\pi/3}\\
\sqrt{3}&-\sqrt{3}&0&0
\end{array}
\right) .
\eeq
It contains the transformation $\tilde{s}_{i,A}$ between the simple objects $i \in \{1,2,3,\sigma\}=\mathcal{C}$ of the input category and the fluxons $A\in \{\mathbf{1},2,7,8\}=\mathcal{F}$ of the output category $\mathcal{Z(C)}$. 
%As in the previous example, the first row is $\tilde{s}_{1,A}=\sqrt{d_A/\mathcal{D}}$ and the first column $\tilde{s}_{i,\mathbf{1}}=d_i/\sqrt{\mathcal{D}}$.

%
%
%%%%%%%%%%%%%%%%%%%
\subsection{Haagerup category $\mathcal{H}_3$}
\label{sec:haa}
%%%%%%%%%%%%%%%%%%%
%
%

The Haagerup category $\mathcal{H}_3$ is a good example of the universality of our formula: it is neither commutative, nor braided, nor Abelian, nor does it respect tetrahedral symmetry, see e.g.~\cite{Hong08}. It has $N_\mathcal{C}=6$ simple objects $\lbrace 1, \alpha, \alpha^*, \rho, \alpha\rho, \alpha^*\rho \rbrace$ with quantum dimensions $\lbrace 1, 1, 1, d_{\rho}, d_{\rho}, d_{\rho} \rbrace$ where $d_{\rho}=\frac{3+\sqrt{13}}{2}$. The Drinfeld center $\mathcal{H}_3$ contains $N_\mathcal{Z}=12$ simple objects $\{\mathbf{1},\mu^1,\mu^2,\mu^3,\mu^4,\mu^5,\mu^6,\pi_1,\pi_2,\sigma^1,\sigma^2,\sigma^3\}$ with quantum  dimensions $\{1, 3d_{\rho}, 3d_{\rho}, 3d_{\rho} , 3d_{\rho} , 3d_{\rho}, 3d_{\rho} ,3 d_{\rho} +1 , 3d_{\rho}+2, 3d_{\rho}+2 ,3d_{\rho}+2 ,3d_{\rho}+2 \}$ so that the total quantum dimension is $\mathcal{D}=3(1+d_\rho^2)$~\cite{Hong08}. The internal multiplicities are given by~\cite{Vanhove22}
%
%
%%%%%%%%%%%%%
\beq
n_{\mathcal{H}_3}=\left(
\begin{tabular}{cccccc}
1&0&0&0&0&0\\
0&0&0&1&1&1\\
0&0&0&1&1&1\\
0&0&0&1&1&1\\
0&0&0&1&1&1\\
0&0&0&1&1&1\\
0&0&0&1&1&1\\
1&0&0&1&1&1\\
2&0&0&1&1&1\\
0&1&1&1&1&1\\
0&1&1&1&1&1\\
0&1&1&1&1&1
\end{tabular}
\right),
\eeq 
%%%%%%%%%%%%%
%
%
so that the three fluxons are $\mathbf{1}$, $\pi_1$, and $\pi_2$. Some degeneracies computed from Eq.~(\ref{eq:result3}) are given in Table~\ref{tab:h3}.
\begin{table}[h!]
\center
\begin{tabular}{| c || c | c | c | c |}
\hline
$D_{\mathcal{H}_3}$ & $(g,b)=(0,0)$ & $(0,1)$ & $(0,2)$ & $(1,0)$   \\ \hline \hline
$q=0$ & 1 & 1 & 6 & 12  \\ \hline   
1 & 0 & 5 & 57 &  33 \\ \hline 
2 & 5 & 52 & 1311 &  1245 \\ \hline   
3 & 47 & 1259 & 42384 &   42000\\ \hline      
4 & 1212 & 41125 & 1456539 & 1454673 \\ \hline 
\end{tabular}
\caption{Degeneracies of the $q^{th}$ excited state of a string net built from  $\mathcal{H}_3$   for various surface topologies up to $q=4$. Here $g$ is the genus and $b$ is the number of boundaries. }
\label{tab:h3}
\end{table}

The Hilbert-space dimension [see Eq.~(\ref{eq:totaldegenresult})] is given by
%
%
%%%%%%%%%%%%%
\beq
\dim \mathcal{H} = \left[1+\frac{1}{(3d_\rho + 1)^{\frac{N_\text{v}}{2}}}+\frac{2^{2-2g+\frac{N_\text{v}}{2}}}{(3d_\rho + 2)^{\frac{N_\text{v}}{2}}} \right] \mathcal{D}^{\frac{N_\text{v}}{2}},
\eeq 
%%%%%%%%%%%%%
%
%
(see also Table~\ref{hilbertdim}).

%
%
%%%%%%%%%%%%%%%%%%%
\subsection{Hagge-Hong category $\mathcal{E}$}
\label{sec:hh}
%%%%%%%%%%%%%%%%%%%
%
%
This is a simple example of an input category $\mathcal{C}=\mathcal{E}$ with fusion multiplicities~\cite{Hong08}. It is commutative, not braided and it breaks the tetrahedral symmetry. It contains $N_\mathcal{C}=3$ simple objets $\{1,x,y\}$ with quantum dimensions $\{1,d_x,1\}$, where $d_x=\sqrt{3}+1$ (see Ref.~\cite{Hong08} for more details). 
%The nontrivial fusion rules are $y\times y=1$, $x\times y = x$ and $x\times x = 1+2 x + y$ and the quantum dimensions are $d_s=\{1,1,d_x\}$, where $d_x=\sqrt{3}+1$. The $F$-symbols are given in Ref.~\cite{Hong08}. 
The Drinfeld center $\mathcal{Z}(\mathcal{E})$ contains $N_\mathcal{Z}=10$ simple objects  $\{\mathbf{1},Y,X_1,X_2,X_3,X_4,X_5,U,V,W\}$ with quantum dimensions $\{1,1,d_x, d_x, d_x, d_x, d_x, d_x +1, d_x +1,d_x +2\}$ so that $\mathcal{D}=2d_x+4$. 
Internal multiplicities are given by~\cite{Hong08}
%
%
%%%%%%%%%%%%%
\beq
n_\mathcal{E}=\left(
\begin{tabular}{ccc}
1&0&0\\
0&0&1\\
0&1&0\\
0&1&0\\
0&1&0\\
0&1&0\\
0&1&0\\
1&1&0\\
0&1&1\\
1&1&1
\end{tabular}
\right),
\eeq 
%%%%%%%%%%%%%
%
%
so that the three fluxons are $\mathbf{1}$, $U$, and $W$. This allows us to compute the degeneracies for a few systems (see Table~\ref{tab:E}).
%
%
%%%%%%%%%
\begin{table}[h!]
\center
\begin{tabular}{| c || c | c | c | c |}
\hline
$D_{\mathcal{E}}$ & $(g,b)=(0,0)$ & $(0,1)$ & $(0,2)$ & $(1,0)$   \\ \hline \hline
$q=0$ & 1 & 1 & 3 & 10  \\ \hline   
1 & 0 & 2 & 11 &  4 \\ \hline 
2 & 2 & 9 & 75 &  82 \\ \hline   
3 & 7 & 66 & 611 &   604\\ \hline      
4 & 59 & 545 & 5139 & 5146 \\ \hline 
\end{tabular}
\caption{Degeneracies of the $q^{th}$ excited state of a string net built from  $\mathcal{E}$   for various surface topologies up to $q=4$. Here $g$ is the genus and $b$ is the number of boundaries. }
\label{tab:E}
\end{table} 
%%%%%%%%%
%
%
The Hilbert-space dimension [see Eq.~(\ref{eq:totaldegenresult}] is therefore
%
%
%%%%%%%%%%%%%
\beq
\dim \mathcal{H} = \left[1+(\sqrt{3}+2)^{\frac{-N_\text{v}}{2}} + (\sqrt{3}+3)^{\frac{-N_\text{v}}{2}} \right] \mathcal{D}^{\frac{N_\text{v}}{2}},
\eeq 
%%%%%%%%%%%%%
%
%
(see  Table~\ref{hilbertdim}). %More generally, we checked that the formula for the degeneracy (\ref{eq:result2}) and the fluxon identity (\ref{eq:identityS}) hold in the case of a fusion multiplicity $>1$. 
%However, we did not compute the degeneracy numerically in the link basis on small clusters.

The mock $S$-matrix is
\beq
\tilde{s}=\frac{1}{\sqrt{2d_x+4}}\left(
\begin{array}{ccc}
1&\sqrt{d_x+1}&\sqrt{d_x+2}\\
d_x&-\sqrt{2}&0\\
1&\sqrt{d_x+1}&-\sqrt{d_x+2}
\end{array}
\right) ,
\eeq
where $\tilde{s}_{i,A}$ with $i \in \{1,x,y\}$ and $A\in \{\mathbf{1},U,W\}=\mathcal{F}$. 
%Again, it was chosen such that the first row is $\tilde{s}_{1,A}=\sqrt{d_A/\mathcal{D}}$ and the first column $\tilde{s}_{i,\mathbf{1}}=d_i/\sqrt{\mathcal{D}}$.

%
%
%%%%%%%%%%%%%%%%%%%
%%%%%%%%%%%%%%%%%%%
\section{Conclusion and perspectives}
\label{sec:conclusion}
%%%%%%%%%%%%%%%%%%%
%%%%%%%%%%%%%%%%%%%
%
%

In this work, we have studied the generalized string-net model with arbitrary input category and restricted to the charge-free sector (all vertex constraints are satisfied) so that only fluxons (plaquette excitations) are present as real excitations at the single plaquette level. This does not exclude the presence of other excitations resulting from the fusion of several fluxons. The main results are analytical expressions for the level degeneracies. In particular, in the case of a noncommutative input category, we show that part of the degeneracy is contained in nontrivial factors $n_{A,1}$ obtained from the tube algebra and not  only from the output category (Drinfeld center). An important result is the generalization of the Moore-Seiberg-Banks formula, Eq.~(\ref{eq:degeq}), that includes this nontopological degeneracy. For example, two Morita-equivalent categories such as Rep$(S_3)$ and Vec$(S_3)$ have the same Drinfeld center $\mathcal{Z}(S_3)$ but the corresponding string-net models have different fluxons and different Hilbert spaces.
In a forthcoming publication~\cite{Ritz23b}, we will use these degeneracies to compute the partition function of the generalized string-net model and to study its finite temperature properties. 

Finally, we wish to use a similar approach to extend the computation of level degeneracies to models (such as Kitaev quantum double~\cite{Kitaev03} or extended string-nets~\cite{Hu18}) that have all elementary excitations of the Drinfeld center and not only fluxons. 

\bigskip

\acknowledgements
We thank N. Bultinck, B. Dou\c{c}ot, S. Dusuel, L. Lootens, and J. Slingerland for illuminating discussions. A.R.-Z. and J.-N.F. acknowledge financial support from  iMAT Sorbonne Universit\'e Emergence through a ``bourse exploratoire''. J.-N.F. acknowledges the support of  the quantum information center at Sorbonne Universit\'e (QICS) via an Emergence grant for the TQCNAA project.  S.H.S. acknowledges support from EPSRC Grant
EP/S020527/1. Statement of compliance with EPSRC
policy framework on research data: This publication is
theoretical work that does not require supporting research data.

\bigskip

\appendix

%
%
%%%%%%%%%%%%%%%%%%%
%%%%%%%%%%%%%%%%%%%
\section{Summary of the tube algebra}
\label{app:tube}
%%%%%%%%%%%%%%%%%%%
%%%%%%%%%%%%%%%%%%%
%
%

This appendix contains an introduction to the tube algebra $\mathcal{TA}$ of a given UFC $\mathcal{C}$ and to its decomposition, as a way of finding the topological quasiparticles, i.e.,  the simple objects of the Drinfeld center $\mathcal{Z}(\mathcal{C})$. The tube algebra was introduced by Ocneanu~\cite{Ocneanu94,Ocneanu01}. A detailed account is given by Lan and Wen~\cite{Lan14}, who call it the $Q$-algebra, which we mainly follow. For a pedagogical introduction, see Ref.~\cite{SimonBook}. 

\begin{figure}[h!]
\includegraphics[width=5cm]{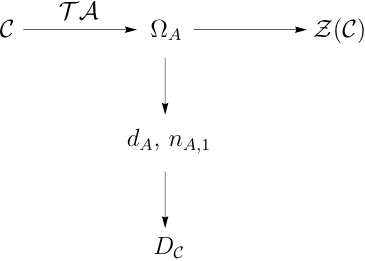}
\caption{Decomposing the tube algebra $\mathcal{TA}$ gives the half-braidings $\Omega_A$ (or, almost equivalently, the minimal central projectors $P_A$) from the input category $\mathcal{C}$. From there, the Drinfeld center $\mathcal{Z}(\mathcal{C})$ follows, as well as the internal multiplicities $n_{A,1}$ and the quantum dimensions $d_A$, that give access to the energy-level degeneracies $D_\mathcal{C}$ of the string-net model.}
\label{fig:taflowdiag}
\end{figure}
Objects in $\mathcal{TA}$ are called tubes and denoted $Q$, following Ref.~\cite{Lan14}. The composition rules of these objects constitute the essence of the tube algebra. From these composition rules, one builds a canonical representation of $\mathcal{TA}$. This representation can be block-diagonalized and the irreducible blocks correspond to the topological sectors. Decomposing the tube algebra means finding its center $\mathcal{Z}(\mathcal{TA})$, containing objects in $\mathcal{TA}$ that commute with every tube. In other words, it means obtaining the projectors $P_A$ onto the topological quasiparticles $A \in \mathcal{Z}(\mathcal{C})$ as linear combinations of the tubes $Q$. Almost equivalently, it means obtaining the half-braidings $\Omega_A$, that appear in the standard definition of the Drinfeld center $\mathcal{Z}(\mathcal{C})$. 

From the half-braidings, it is easy to obtain all of the properties of the Drinfeld center, e.g., the $S$ and $T$ matrices, and therefore the fusion matrices, the quantum dimensions of the simple objects~\cite{Levin05,Lan14}), and more informations such as the internal multiplicities. The multiplicities $n_{A,1}$ and the quantum dimensions $d_A$ of the quasiparticles $A \in \mathcal{Z}(\mathcal{C})$ are all that is needed to compute the degeneracies of the energy levels of a string-net model [see Eq.~(\ref{eq:gsqclosed}) and Fig.~\ref{fig:taflowdiag}]. 

In the following, we first define the tubes and obtain their algebra. Then, we explain how to decompose the tube algebra in order to find the topological quasiparticles and their internal multiplicities.

\subsection{Tubes $Q$ and their algebra $\mathcal{TA}$}
\begin{figure}[h!]
\includegraphics[width=6.5cm]{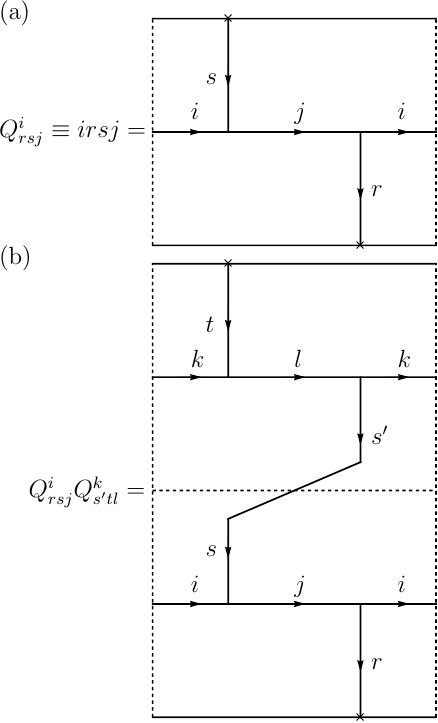}
\caption{(a) Graphical representation of a tube $Q_{rsj}^i$. The vertical dashed lines indicate periodic boundary conditions so that the rectangle is topologically equivalent to a tube or cylinder. The crosses indicate the open ends of the vertical strings. (b) Vertical stacking of two tubes $Q^i_{rsj} Q^k_{s' t l} $ The horizontal dashed line indicates the stacking.\label{fig:tubeQ}}
\end{figure}

The main objects of $\mathcal{TA}$ are the tubes $Q_{rsj}^i\equiv irsj $ [see Fig.~\ref{fig:tubeQ}(a)], where $i,r,s,j$ are simple objects of $\mathcal{C}$. A tube has two open string ends, labeled by $r$ and $s$, and two strings without open ends called $i$ and $j$. Here the vertex indices are dropped.
%
%
%%%%%%%%%%%%%
%%%%%%%%%%%%%
%
% 

The only allowed tubes are those that respect the fusion rules of $\mathcal{C}$, i.e.,  $i\times j = \sum_k N_{ij}^k \, k$ at their two inner vertices. Therefore, the number of tubes $N_\mathcal{T}$, i.e. the dimension of the tube algebra, is given by a product of two fusion matrices (applying to the two vertices):
%
%
%%%%%%%%%%%%%
\beq
%\dim \mathcal{T A} = 
N_\mathcal{T}= \sum_{i,j,r,s} N_{s i}^j N_{j \bar r}^i .
\label{eq:nt}
\eeq 
%%%%%%%%%%%%%
%
%
The tubes can be composed (or multiplied):  $Q^i_{rsj}Q^k_{s' t l}$ means that $Q^k_{s' t l}$ is stacked on top of $Q^i_{rsj}$ [see Fig.~\ref{fig:tubeQ}(b)]. The $\mathcal{TA}$ has the following structure~\cite{Lan14}
%
%
%%%%%%%%%%%%%
\beq\label{eq:ta}
Q^i_{rsj} Q^k_{s' t l} =\sum_{m,n} \delta_{ss'}\sqrt{\frac{d_i d_k}{d_m}} F^{i \overline{j} s}_{\overline{k} l \overline{n}} F^{\overline{r} \overline{i} j}_{\overline{k} n \overline{m}} F^{t k \overline{l}}_{i \overline{n} m} \, \, Q^m_{rtn} ,
\eeq 
%%%%%%%%%%%%%
%
%
and is noncommutative in general. The coefficients %$\delta \sqrt{...} F F F$ 
in front of $Q^m_{rtn}$ in the sum on the right-hand side are the structure constants of the tube algebra. 
One can observe that the stacking is determined by the labels $r$ and $s$ of the open strings. Therefore, the tube algebra splits in different sectors labeled by $rs$. This will play a role later in the decomposition of the tube algebra.

Two important sets of tubes deserve a particular attention:
%
%
%%%%%%%%%%%%%
\begin{figure}[h!]
\includegraphics[width=6.5cm]{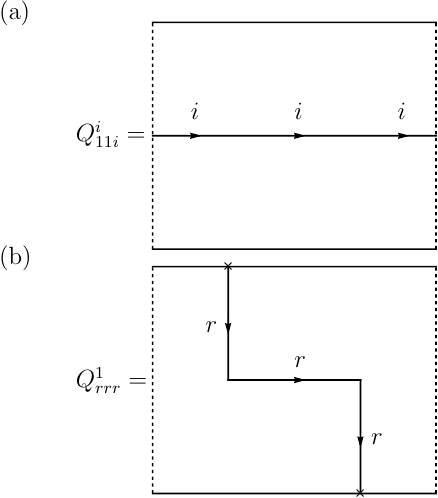}
\caption{Two particular sets of tubes: (a) horizontal closed string $Q_{11i}^i$; (b) vertical open string $Q^1_{rrr}$. %behaving as $\mathds{1}_{rr}$.
} \label{fig:tubeQ2}
\end{figure}
%%%%%%%%%%%%%
%
% 
(1) the $N_\mathcal{C}$ tubes $Q_{11i}^i$ which have no open ends and are formed by ``horizontal'' closed strings  [see Fig.~\ref{fig:tubeQ2}(a)]. Each such tube corresponds to a simple object $i$ of $\mathcal{C}$. The $11$ sector consists only of these tubes and the tube algebra restricted to the $11$ sector reflects directly the algebra of the input category:
\beq
Q_{11i}^i Q_{11j}^j = \sum_k N_{i j}^k Q_{11k}^k .
\label{eq:qqq}
\eeq

(2) The $N_\mathcal{C}$ tubes $Q^1_{rrr}$ formed by a ``vertical'' open string [see Fig.~\ref{fig:tubeQ2}(b)] which correspond to the identity operator $\mathds{1}_{rr}$ in the $rr$ sector of $\mathcal{TA}$:
%
%
%%%%%%%%%%%%%
\beq\label{eq:1rrr}
Q^1_{rrr} Q^k_{r r l} = Q^k_{r r l}Q^1_{rrr}= Q^k_{r r l}.
\eeq 
%%%%%%%%%%%%%
%
%
Graphically, it is easy to see that stacking $Q^1_{rrr}$ on top or below of any tube of the same sector returns that second tube. The identity over $\mathcal{TA}$ is the sum over all of the $\mathds{1}_{rr}$: 
%
%
%%%%%%%%%%%%%%%%%%
\beq\label{eq:idta}
\mathds{1} = \sum_r \mathds{1}_{rr} = \sum_r Q^1_{rrr}.
\eeq
%%%%%%%%%%%%%%%%%
%
%

\subsection{Decomposition of the tube algebra}
The Wedderburn-Artin theorem states that a semi-simple algebra (such as $\mathcal{TA}$) is isomorphic to a direct sum of simple matrix algebras (see, e.g., Ref.~\cite{Lootens22}). It means that the tube algebra splits into $N_\mathcal{Z}$ matrix algebras $\mathcal{M}_n$ of dimension $n$: 
\beq 
%\mathcal{TA} \simeq \mathcal{M}_{n_1}\oplus \mathcal{M}_{n_2}\oplus ...\oplus \mathcal{M}_{n_{N_\mathcal{Z}}}
\mathcal{TA} \simeq \oplus_{n_A} \mathcal{M}_{n_A},
\eeq
with $A=1,...,N_\mathcal{Z}$. Each block corresponds to an irreducible representation (or simple module) of the algebra and therefore to a simple object (a topological quasiparticle) $A$ in the Drinfeld center $\mathcal{Z}(\mathcal{C})$.

In the following, we represent the tube algebra in a vector space and search for its block structure in order to identify the irreducible representations $A$ and their dimensions $n_A$.

\subsection{Simple idempotents and nilpotents $p$}
A first step is to construct a vector space of dimension $V=\sum_A n_A$ in which to represent the tube algebra by finding an orthonormal basis of vectors $\{|\alpha \rangle\}$. When decomposing the tube algebra, one does not know \textit{a priori} the dimension $V$ of the vector space, the labels $A$ of the irreducible representations, their dimensions $n_A$ or their total number $N_\mathcal{Z}$. These are the output of the decomposition. Therefore, for the moment, we do not give the final labels of these states and just use greek letters such as $\alpha$ as a generic label (later, we will see that $\alpha$ can actually correspond to several labels).

The basis is found by building $V$ simple orthogonal idempotents (or projectors) $p$ as linear combinations of the tubes $Q$. Simple means that they cannot be further decomposed or, in other words, that they are projectors onto one-dimensional subspaces. We denote simple idempotents by $p^{\alpha \alpha}=|\alpha \rangle \langle \alpha|$ (known as ``diagonal elements'', as later we will also introduce ``off-diagonal elements'' like $p^{\alpha \beta}$ with $\beta\neq \alpha$) and they satisfy the orthogonality relation
%
%
%%%%%%%%%%%%%
\beq
p^{\alpha \alpha} p^{\beta \beta} = \delta_{\alpha,\beta}\,  p^{\alpha \alpha},
\eeq 
%%%%%%%%%%%%%
%
%
and $\tr\, p^{\alpha \alpha} =1$, meaning that the projected subspace is one dimensional. The identity in the vector space is given by the sum over all these simple idempotents:
\beq 
\mathds{1}=\sum_{\alpha} p^{\alpha \alpha}.
\eeq

The next step is to find which simple idempotents belong to the same irreducible block or, in other words, which simple idempotents correspond to the same topological quasiparticle $A$. In order to do so, we need to find $(N_\mathcal{T}-V)$ nilpotents $p^{\alpha \beta}$ with $\alpha\neq \beta$, which are also linear combinations of the tubes that satisfy $(p^{\alpha \beta})^2=0$. They are traceless and connect one-dimensional subspaces that belong to the same block. For example, if 
%
%
%%%%%%%%%%%%%
\beq
p^{\alpha \beta} p^{\beta \alpha} = p^{\alpha \alpha}, \text{ and } p^{\beta \alpha} p^{\alpha \beta}  = p^{\beta \beta},
\eeq 
%%%%%%%%%%%%%
%
%
it means that $|\alpha\rangle$ and $|\beta\rangle$ belong to the same irreducible block (say $A$) and that $p^{\alpha \beta}=|\alpha\rangle \langle \beta|=(p^{\beta\alpha})^\dagger$ (``off-diagonal elements''). Therefore, we denote $p_A^{\alpha \alpha}$, $p_A^{\beta \beta}$, $p_A^{\alpha \beta}$ and $p_A^{\beta \alpha}$ the simple idempotents and nilpotents that belong to the same irreducible block. 

More generally, simple idempotents and nilpotents satisfy the following key equation:
%
%
%%%%%%%%%%%%%
\beq
p_A^{\alpha \beta} p_B^{\gamma \delta} = \delta_{A,B}\delta_{\beta,\gamma} p_A^{\alpha \delta}.
\label{eq:key}
\eeq 
%%%%%%%%%%%%%
%
%
In total, there are as many simple idempotents and nilpotents as tubes. When we want to refer to them collectively, we will call them ``idemnils'' and use a small $p$.

\subsection{Minimal central idempotents $P_A$}
Once we have found the $N_\mathcal{T}$ idemnils and we have identified the ones that belong to the same irreducible block, we construct the minimal (or simple) central idempotents $P_A$. Similarly to the simple idempotents, they are orthogonal idempotents
%
%
%%%%%%%%%%%%%
\beq
P_A P_B = \delta_{A,B} P_A,
\eeq 
%%%%%%%%%%%%%
%
%
but they also commute with any object of $\mathcal{TA}$, i.e., they belong to the center $\mathcal{Z}(\mathcal{TA})$ of the tube algebra. In addition, they are minimal, i.e., they cannot be decomposed further into a sum of other central idempotents. By summing over all simple idempotents that belong to the same irreducible block, we get 
%
%
%%%%%%%%%%%%%
\beq
P_A = \sum_{\alpha} p_A^{\alpha \alpha},
\label{eq:PA}
\eeq 
%%%%%%%%%%%%%
%
%
which is the projector onto a topological quasiparticle $A$. The dimension of the irreducible representation $A$ (or the multiplicity of $A$) is 
%
%
%%%%%%%%%%%%%
\beq
n_A=\tr \, P_A.
\eeq 
%%%%%%%%%%%%%
%
%
We can also define partial multiplicities such as
%
%
%%%%%%%%%%%%%
\beq
n_{A,\alpha}=\tr \, p_A^{\alpha \alpha}= 0 \text{ or }1,
\label{eq:nAalpha}
\eeq 
%%%%%%%%%%%%%
%
%
for the topological quasiparticle $A$ in the $\alpha$ sector. Here it is important to realize that $\alpha$ may denote either a single label (this will be the case for a commutative input category) or a pair of labels (this will be the case for a noncommutative input category).

The number of minimal central idempotents (topological sectors), $N_\mathcal{Z}$ is such that:
%
%
%%%%%%%%%%%%%
\beq
N_\mathcal{Z} = \sum_A 1 \leqslant V =\sum_A n_A \leqslant  N_\mathcal{T} = \sum_A n_A^2 .
\eeq 
%%%%%%%%%%%%%
%
%
%For other properties of the minimal central idempotents, see Appendix~\ref{app:stringoperators}. 

\subsection{Tube algebra sectors}
When decomposing the tube algebra, a substantial simplification comes from the fact that, as we have seen, the tube algebra separates into independent sectors labeled by the open ends $r$ and $s$ of a tube $Q^i_{rsj}$. Therefore, the search for idemnils can be done sector by sector (indexed by $r s$). A simple idempotent necessarily belongs to a diagonal sector $rr$, whereas a nilpotent belongs to an off-diagonal sector $rs$ with $s\neq r$.

If the input category $\mathcal{C}$ is commutative, the tube algebra in a given sector $rs$ is commutative so that the corresponding vector space has a dimension $V^{rs}$ equal to the number of tubes 
%
%
%%%%%%%%%%%%%
\beq
N_\mathcal{T}^{rs}= \sum_{i,j} N_{s i}^j N_{j \bar r}^i.
\label{eq:ntrs}
\eeq 
%%%%%%%%%%%%%
%
%
By summing over all sectors $rs$, one recovers Eq.~(\ref{eq:nt}).

A simple idempotent is labeled as $p_A^{rr}$ and a nilpotent as $p_A^{rs}$ with $s\neq r$ (meaning that $\alpha$ is actually $r \in \mathcal{C}$ here).  Also, one has that
%
%
%%%%%%%%%%%%%
\beq
n_{A,r} = \tr \, p_A^{rr} = 0 \text{ or }1.
\eeq 
%%%%%%%%%%%%%
%
%
If the input category $\mathcal{C}$ is noncommutative, the tube algebra in a given $rs$ sector need not be commutative. If it is noncommutative, it means that the corresponding vector space's dimension $V^{rs}<N_\mathcal{T}^{rs}$ and that one needs to introduce an extra index $a$ to label orthonormal basis vectors. More generally, a simple idempotent is labeled as $p_A^{rr,aa}$ and a nilpotent as $p_A^{rs,ab}$ with $s\neq r$ or $b\neq a$ (meaning that $\alpha$ is actually $r,a$ with $r\in \mathcal{C}$ and $a=1,...,n_{A,r}$ here). We call $P_A^{rr}=\sum_a p_A^{rr,aa}$ the projector onto the quasiparticle $A$ in sector $rr$ (our convention is that an idempotent denoted by a small $p$ is a simple idempotent that has a trace of $1$, whereas a capital $P$ is a projector with a trace that can be larger than $1$. Also note that $P_A^{rr}$, in contrast to $P_A$, does generally not belong to the center of the tube algebra). It projects onto a subspace of dimension
%
%
%%%%%%%%%%%%%
\beq
n_{A,r} = \tr \, P_A^{rr},
\eeq 
%%%%%%%%%%%%%
%
%
which can also be interpreted as a multiplicity for the topological quasiparticle $A$ in a given sector. An important fact is that, for a noncommutative input category, $n_{A,r}$ is no longer restricted to $0$ or $1$ but can be larger. Examples are given in Sec.~\ref{sec:vecrep} and in Sec.~\ref{sec:haa}.

The eigenvectors $\lbrace \ket{A,r,a} \rbrace$ (of eigenvalue $1$) of the simple idempotents form an orthonormal basis of the total vector space. The idemnils can be written as
\beq
p_A^{rs, ab}= \ket{A,r,a}\bra{A,s,b},   
\eeq 
and Eq.~(\ref{eq:key}) becomes
%
%
%%%%%%%%%%%%%
\beq
p_A^{rs,aa'} p_B^{s' t,b'b} = \delta_{A,B} \delta_{s,s'}\delta_{b',a'}p_A^{rt,ab}.
\eeq 
%%%%%%%%%%%%%
%
%
%
Equation~(\ref{eq:nAalpha}) becomes $n_{A,r,a}=\tr \, p_A^{rr,aa}=0$ or $1$ even if $n_{A,r}$ can be strictly greater than $1$.

Next, each $P_A$ is constructed as a sum of the corresponding simple idempotents $p_A$, so that Eq.~(\ref{eq:PA}) becomes:
%
%%%%%%%%%%%%%
\beq
P_A = \sum_{r \in \mathcal{C}} P_A^{rr}
= \sum_{r \in \mathcal{C}} \sum_{a=1}^{n_{A,r}} p_A^{rr,aa}.
\eeq 
%%%%%%%%%%%%%
%
%
The multiplicity of $A$ (or the dimension of the irreducible representation $A$) is given by: 
\beq
    n_A = \sum_r n_{A,r}.
\eeq

\subsection{The $11$ sector and fluxons}
\label{sec:fluxons}
When considering generalized string-net models in the charge-free sector, one particular type of tubes plays a major role: the tubes in the $11$ sector. These $N_\mathcal{C}$ tubes are formed by closed horizontal strings [see Fig.~\ref{fig:tubeQ2}(a)] and have no open ends. Because of this latter characteristic, they correspond to pure plaquette excitations, i.e., fluxons. Among the objects of $\mathcal{Z}(\mathcal{C})$, fluxons are identified through the fact that the corresponding $P_A$ has non-zero weight on the $11$ sector, i.e.,
\beq
n_{A,1}\geqslant 1 .
\label{eq:nalpha1}
\eeq 
In the $11$ sector, the number of tubes $N_\mathcal{T}^{11}=N_\mathcal{C}$ is larger or equal  than the vector space dimension $V^{11}$ which is larger or equal  than the number of fluxons $N_\mathcal{F}$:
%
%
%%%%%%%%%%%%%
\beq
N_\mathcal{C} = \sum_A n_{A,1}^2 \geqslant V_{11} =\sum_A n_{A,1} \geqslant N_\mathcal{F} = \sum_A \text{sgn } n_{A,1} .
\eeq 
%%%%%%%%%%%%%
%
%
In the above equation, the equalities occur when $\mathcal{C}$ is commutative, i.e. when $n_{A,1}$ is either $0$ or $1$. Moreover, the projector on the vacuum particle $A=\mathbf{1}$ of $\mathcal{Z}(\mathcal{C})$ is a weighted sum of all tubes in the $11$ sector: 
\beq 
P_\mathbf{1} = p_\mathbf{1}^{11} = \sum_{i} \frac{d_i}{\mathcal{D}} Q^{i}_{11i}.
\label{eq:projvacuumqp}
\eeq 
This projector is equal to the ``Kirby strand"~\cite{SimonBook}. When the contour over which it acts is a single plaquette, $P_\mathbf{1}$ is the same as $B_p$ defined in Sec.~\ref{sub:Hamiltonian}. 
%See also the discussion of this projector in Appendix~\ref{app:stringoperators}.

\subsection{From tubes to idemnils: $M_A$}
Finally, we express the tubes as linear combinations of the idemnils (this corresponds to writing the irreducible representations of the tubes in the basis spanned by the vectors $\ket{A, r, a}$) : 
%
%
%%%%%%%%%%%%%
\beq
Q_{rsj}^i =
%\sum_{A, r', s', a, b} M_{i r s j, A r' s' a b}\,  p_A^{r's',ab} \nn \\
\sum_{A, a, b} (M^i_{A, rsj})_{a,b} \,  p_A^{rs,ab},
\label{eq:qmp}
\eeq 
%%%%%%%%%%%%%
%
%
where
%
%
%%%%%%%%%%%%%
\beq
(M^i_{A, rsj})_{a,b} = \langle A, r, a |Q_{rsj}^i |A,s,b\rangle.
\label{eq:MQAsb}
\eeq 
%%%%%%%%%%%%%
%
%
The coefficients $M_A$ in this decomposition are sometimes called modules (see Ref.~\cite{Lan14}).

The quantity $M^i_{A, rsj}$ is often written as an $N_\mathcal{Z}\times N_\mathcal{T}$ matrix with row index $A$ and column index $irsj$. Each element $(M^i_{A, rsj})_{a,b}$ is itself a small matrix with row index $a=1,\dots,n_{A,r}$ and column index $b=1,\dots,n_{A,s}$. 

The internal multiplicities (and in particular the important $n_{A,1}$) are given by:
\beq
n_{A,s}=\tr \, M_{A,sss}^1. 
\label{eq:nalphas}
\eeq 
From Eq.~(\ref{eq:nalphas}) and the knowledge of $d_s$, one also obtains the quantum dimensions of the simple objects in $\mathcal{Z}(\mathcal{C})$ as
\beq
d_A=\sum_s n_{A,s} d_s \geqslant n_{A} \geqslant n_{A,1},
\eeq 
where, for the inequalities, we used that $d_s\geqslant 1$ and $n_{A,s}\geqslant 0$. The multiplicity $n_{A,1}$ and the quantum dimensions $d_A$ are all that is needed to compute the spectrum degeneracies [see Eq.~(\ref{eq:result2})]. As a consequence of the above inequality and the fact that $d_\mathbf{1}=1$, when $A=\mathbf{1}$, one has $n_\mathbf{1}=n_{\mathbf{1},1}=1$.

\subsection{Fluxon degeneracies in the Levin-Wen model}
\label{sub:FluxDegLW}
The generalized Levin-Wen model has the property that the degeneracy of states is invariant under restructuring of the underlying graph by local (so-called ``$F$"-) moves that preserve the numbers of vertices, edges, and plaquettes.   In particular this means that to determine the number of states available by a plaquette, we can use the simplest plaquette  possible, which is a single loop, connected to the rest of the graph by a single stem.  The space of states available is then just the tube basis states $Q^i_{r1i}$ which has $i$ going around the loop and the stem labeled with $r$.  We then decompose these basis states into idemnils via Eqs.~(\ref{eq:qmp}) and (\ref{eq:MQAsb}).  Since we are excluding any vertex violations we consider values of $A$ which are fluxons only.  For a particular value of $A$, the basis then consists of 
the orthonormal idemnils $p_A^{r1,ab}$ with $a=1, \ldots, n_{A,r}$ and $b=1, \ldots, n_{A,1}$.  We can think of the factor $n_{A,1}$ as coming from the degeneracy associated with the ``loop" end of the tube.
%, and the degeneracy $n_{A,r}$ is from the end with the $r$-stem.  

\subsection{Half-braidings $\Omega_A$}
Levin and Wen~\cite{Levin05} have developed an alternative way of finding the topological quasiparticles from the input category. Instead of the projectors $P_A$ on the topological quasiparticles, they find the (closed) string operators $W_A$ that also belong to the tube algebra center but obey the fusion algebra of $\mathcal{Z}(\mathcal{C})$:
%
%
%%%%%%%%%%%%%
\beq
W_A W_B = \sum_C N_{A,B}^C W_C.
\label{eq:WW}
\eeq
%%%%%%%%%%%%%
%
%
It is a simple matter to go from the projectors $P_A$'s to the string operators $W_A$'s or vice-versa. For example
%
%
%%%%%%%%%%%%%
\beq
P_A=S_{1,A}\sum_B S_{A,B}^* W_B .
\label{eq:PW}
\eeq
%%%%%%%%%%%%%
%
%

The output of their approach is not $M_A$ but is called $\Omega_A$, which is a half-braiding in the standard construction of the Drinfeld center (see, e.g., Ref.~\cite{Hong08,Wang_book}). Actually the two quantities $M_A$ and $\Omega_A$ are closely related [see Eq.~(62) in Ref.~\cite{Lan14}]:
%
%
%%%%%%%%%%%%%
\beq
M_{A,rsj}^i = \Omega_{A,rsj}^i d_i \sqrt{\frac{d_s}{d_r}}.
\label{MOmega}
\eeq 
%%%%%%%%%%%%%
%
%

Eventually, from $\Omega_A$ (or $M_A$), all properties of the Drinfeld center $\mathcal{Z}(\mathcal{C})$, such as the $S$, $T$ and fusion matrices, can also be easily obtained (see Eqs.~(64), (65) and (60) in Ref.~\cite{Lan14}).

In order to obtain the $S-$ and $T-$ matrices of the Drinfeld center as well as the internal multiplicities $n_{A,1}$, it is enough to know the coefficients of the decomposition of the minimal central idempotents $P_A$ on the tubes $Q^i_{rsj}$. Note that the half-braidings $\Omega_A$ contain a little more information as they give the full decomposition of the idemnils $p_A^{rs}$ on the tubes. This extra information is needed for certain observables such as Wegner-Wilson loops (see Ref.~\cite{Ritz23b}).

\medskip

In practice, finding the block structure of the tube algebra is not always an easy task. Lan and Wen give a procedure that works well in simple cases and which they call idempotent decomposition~\cite{Lan14}. Another, more systematic/algorithmic approach to obtain the minimal central idempotents from the tube algebra is described in Appendix C of Ref.~\cite{Bultinck17}.

In the following appendix, we give the result of the decomposition of the tube algebra for two Morita-equivalent categories.

%
%
%%%%%%%%%%%%%%%%%%%
%%%%%%%%%%%%%%%%%%%
\section{Half-braidings for Rep($S_3$) and Vec($S_3$)}
\label{app:halfbraiding}
%%%%%%%%%%%%%%%%%%%
%%%%%%%%%%%%%%%%%%%
%
%

The Drinfeld center for Rep($S_3$) is given, e.g., in Ref.~\cite{Beigi11}. Some details about its tube algebra can be found in an Appendix in Ref.~\cite{Bultinck17} and the half-braidings are derived in Ref.~\cite{Lootens23}. Since all $\Omega^i_{J, rsj}$ are one-dimensional, we write them under the form of a matrix $\Omega$ with $8$ rows indexed by particles \mbox{$J=\lbrace A,B,C,D,E,F,G,H \rbrace$} and $17$ columns indexed by  tubes 
\begin{widetext}
$ irsj=\lbrace 
1111, %0000,  
2112,%1001, 
3113,%2002, 
1222,%0111, 
2221,%1110, 
3223,%2112, 
1333,%0222, 
2333,%1222, 
3331,%2220, 
3313,%2202, 
3133,%2022, 
3213,%2102, 
3123,%2012, 
3233,%2122, 
3323,%2212
3332,
3333
\rbrace :$ %(notations follow Lan and Wen's convention): 
%
%
%%%%%%%%%%%%%
\beq\hspace*{-0.5cm}
\Omega=\left(
\begin{tabular}{ccccccccccccccccc}
 1 & 1 & 1 & 0&0&0&0&0&0&0&0&0&0&0&0&0&0\\
 0 & 0 & 0 & 1 & 1 & -1 & 0&0&0&0&0&0&0&0&0&0&0\\
 0&0&0&0&0&0&1 & -1 & $\frac{1}{2}$ & -$\frac{1}{2}$ & $\frac{1}{\sqrt{2}}$&0&0&0&0&0&0\\
 1 & -1 & 0& 0 & 0 & 0 & 1 & 1 & $\frac{1}{2}$&  $\frac{1}{2}$& 0 &$2^{\frac{1}{4}}$& $\frac{1}{2^{\frac{3}{4}}}$ &0&0&0&0\\
 0&0&0&1&-1&0&1 & 1 & -$\frac{1}{2}$&  -$\frac{1}{2}$&0&0&0&0&0&-$\frac{i}{2^{\frac{3}{4}}}$ &-$i 2^{\frac{1}{4}}$\\
 1 & 1 & -$\frac{1}{2}$ & 1 & 1 &$\frac{1}{2}$&0&0&0&0&0&0&0&$\frac{\sqrt{3}}{2}$ & $\frac{\sqrt{3}}{2}$&0&0\\
 0&0&0&0&0&0 &1& -1 & $-\frac{1+i \sqrt{3}}{4}$ & $\frac{1+i \sqrt{3}}{4}$& $-\frac{1-i \sqrt{3}}{2^{3/2}}$&0&0&0&0&0&0\\
 0 &0&0&0&0&0&1 & -1& $-\frac{1-i \sqrt{3}}{4}$ & $\frac{1-i \sqrt{3}}{4}$ & $-\frac{1+i \sqrt{3}}{2^{3/2}}$ & 0 & 0& 0& 0&0&0
\end{tabular}
\right).
\eeq 
%%%%%%%%%%%%%
%
%
\end{widetext}

For Vec($S_3$), we have decomposed the tube algebra and give here the resulting half-braidings $\Omega_{J,rsj}^i$ as a matrix with $8$ rows indexed by $J$ as before and $36$ columns indexed by the tubes
\begin{widetext}
%
%
%%%%%%%%%%%%%
\begin{eqnarray}
irsj&=&\{
1111,%0000, 
2112,%1001, 
3113,%2002, 
4114,%3003, 
5115,%4004, 
6116,%5005, 
1222,%0111, 
2223,%1112, 
3221,%2110, 
1333,%0222, 
2331,%1220, 
3332,%2221, 
1444,%0333, 
4441,%3330, 
1555,%0444, 
5551,%4440, 
1666,%0555, 
6661,%5550,
\nn \\ 
&&
4235,%3124, 
5236,%4125, 
6234,%5123, 
4326,%3215, 
5324,%4213, 
6325,%5214, 
2456,%1345, 
6452,%5341, 
3546,%2435, 
6543,%5432, 
3465,%2354, 
5463,%4352, 
2645,%1534, 
5642,%4531, 
2564,%1453, 
4562,%3451, 
3654,%2543, 
4653%3542
\nn\}, 
\end{eqnarray}
%%%%%%%%%%%%%
%
% 
where $s=\{1,2,3,4,5,6\}=\{e,\zeta,\zeta^2,\tau,\tau\zeta,\tau\zeta^2\}$. Defining $\omega= {\rm e}^{2 {\rm i} \pi/3}$ and $\omega^*= {\rm e}^{-2 {\rm i} \pi/3}$, the half-braiding matrix reads
%
%
%%%%%%%%%%%%%
\beq
\Omega=\left(
\begin{tabular}{cccccccccccccccccccccccccccccccccccc}
 1&1&1&1&1&1&0&0&0&0&0&0&0&0&0&0&0&0      &0&0&0&0&0&0&0&0&0&0&0&0&0&0&0&0&0&0\\
 1&1&1&-1&-1&-1&0&0&0&0&0&0&0&0&0&0&0&0      &0&0&0&0&0&0&0&0&0&0&0&0&0&0&0&0&0&0\\
 2&-1&-1&0&0&0&0&0&0&0&0&0&0&0&0&0&0&0      &0&0&0&0&0&0&0&0&0&0&0&0&0&0&0&0&0&0\\
 0&0&0&0&0&0&0&0&0&0&0&0&1&1&1&1&1&1      &0&0&0&0&0&0&1&1&1&1&1&1&1&1&1&1&1&1\\
 0&0&0&0&0&0&0&0&0&0&0&0&1&-1&1&-1&1&-1      &0&0&0&0&0&0&1&-1&1&-1&1&-1&1&-1&1&-1&1&-1\\
 0&0&0&0&0&0&1&1&1&1&1&1&0&0&0&0&0&0      &1&1&1&1&1&1&0&0&0&0&0&0&0&0&0&0&0&0\\
 0&0&0&0&0&0&1&$\omega^*$&$\omega$&1&$\omega$&$\omega^*$&0&0&0&0&0&0      &1&$\omega^*$&$\omega$&1&$\omega$&$\omega^*$&0&0&0&0&0&0&0&0&0&0&0&0\\
 0&0&0&0&0&0&1&$\omega$&$\omega^*$&1&$\omega^*$&$\omega$&0&0&0&0&0&0      &1&$\omega$&$\omega^*$&1&$\omega^*$&$\omega$&0&0&0&0&0&0&0&0&0&0&0&0
\end{tabular}
\right).
\label{eq:OmegaVec}
\eeq 
%%%%%%%%%%%%%
%
% 
As $n_{C,1}=2$ [see Eq.~(\ref{eq:nvec})], beware that $\Omega_{C,11j}^i$ is actually a $2\times 2$ matrix of which Eq.~(\ref{eq:OmegaVec}) only gives the trace. 
\end{widetext}

%
%
%%%%%%%%%%%%%%%%%%%
%%%%%%%%%%%%%%%%%%%
\section{Fluxon identities}
\label{app:condensation}
%%%%%%%%%%%%%%%%%%%
%%%%%%%%%%%%%%%%%%%
%
%

In Appendix~\ref{app:tube} on the decomposition of the tube algebra, fluxons are identified as topological quasiparticles $A$ that have $n_{A,1}>0$. One can define a vector $\mathbf{n}_1$ with components $n_{A,1}$ with $A=1,...,N_\mathcal{Z}$ such that it has non-zero entries only when $A$ is a fluxon. Equations~(\ref{eq:identityS}-\ref{eq:identityT}) are extra relations satisfied by the vector $\mathbf{n}_1$ that we call fluxon identities. Similar equations first appeared in the context of anyon condensation~\cite{Neupert16} and gapped boundaries~\cite{Levin13,Lan15}. Equation~(\ref{eq:identityT}) involving the $T$-matrix means that fluxons are necessarily bosons, i.e., they have a trivial twist $\theta_A=1$. When $\mathcal{Z}(\mathcal{C})$ is Abelian, Eq.~(\ref{eq:identityS}) involving the $S$-matrix means that fluxons also have trivial mutual statistics~\cite{Levin13}. When $\mathcal{Z}(\mathcal{C})$ is non-Abelian, the interpretation is less obvious. In the following, we provide a proof of the fluxon identites from the tube algebra, and comment on the relation with anyon condensation.

\subsection{Proof from the tube algebra}
\label{sec:proof}
%{\red In this section, it is not crystal clear that $Q_1=\sum_A p_A^{11}$ (defined in the tube algebra) is the same as the column vector $\mathbf{n}_1$ with components $n_{A,1}$. Same question about the relation between operator $S$ acting on the tubes (generator of the modular group) and the matrix with components $S_{A,B}$ (defined in the UMTC $\mathcal{Z(C)}$).}

The tubes in the $11$ sector are the horizontal closed (input) strings $Q^i_{11i}$. Therefore, the tube algebra restricted to the $11$ sector is just the fusion algebra of the input category and it decouples from the rest of the tube algebra [see Eq.~(\ref{eq:qqq})].
%\beq
%Q^i_{11i} Q^j_{11j} = \sum_k N_{i j}^k Q^k_{11k}.
%\eeq
In particular
\beq
Q^1_{111} Q^1_{111} = \sum_k N_{1 1}^k Q^k_{11k} = Q^1_{111},
\eeq
is the projector onto the $11$ sector (it is the empty tube). The vertical tubes $Q^1_{rrr}$ are the projectors onto the $rr$ sectors [see Eq.~(\ref{eq:1rrr})].

\subsubsection{Commutative input category}
\label{sec:mock}
In the commutative case, we can use the mock $S$-matrix $\tilde{s}$ to diagonalize simultaneously all the fusion matrices $N_i$. It is a unitary matrix but, unlike the $S$-matrix, it is not symmetric in general (a special case is when the input category is modular, in which case $\tilde{s}$ is a genuine $S$-matrix). Naturally, one would label the matrix elements as $\tilde{s}_{i,j}$ with $i$ and $j \in \mathcal{C}$. However, $\tilde{s}$ is also a unitary transformation between input strings and fluxons (see Eqs. (\ref{eq:c2}) and (\ref{eq:c3}) below). As such, it makes physically more sense to write a matrix element of $\tilde{s}$ as $\tilde{s}_{i,A}$ with rows indexed by input labels $i=1,..,N_\mathcal{C}$ and columns by fluxons $A \in \mathcal{F}$ with $N_\mathcal{F}=N_\mathcal{C}$. The matrix $\tilde{s}$ can be chosen such that its first row only contains strictly positive elements, in which case, $\tilde{s}_{1,A}=\sqrt{d_A/\mathcal{D}}=\sqrt{S_{\mathbf{1},A}}$. The mock $S$-matrix is not yet unique as its columns can always be permuted. By convention, we choose that the first column corresponds to the vacuum $A=\mathbf{1}$ of the output category so that $\tilde{s}_{i,\mathbf{1}}=d_i/\sqrt{\mathcal{D}}$. Examples of mock $S$-matrix are given in Secs.~\ref{sec:vecrep}, \ref{sec:ty3} and \ref{sec:hh}. 

In all of Sec.~\ref{sec:proof}, $A$ is taken to be a fluxon. The simple idempotents $p_A^{11}$ are easily found as linear combination~\cite{footnote1} of the horizontal tubes $Q^i_{11i}$
\beq
p_A^{11} = \tilde{s}_{1,A} \sum_i \tilde{s}_{i,A}^* Q^i_{11i},
\label{eq:c2}
\eeq
and vice-versa:
\beq
Q^i_{11i}  =  \sum_A \frac{\tilde{s}_{i,A}}{\tilde{s}_{1,A}}  p_A^{11}.
\label{eq:c3}
\eeq
Using the Verlinde-like equation
\beq
N_{i j}^k =\sum_A \frac{\tilde{s}_{i,A} \tilde{s}_{j,A} \tilde{s}_{k,A}^*}{\tilde{s}_{1,A}},
\label{eq:Verlindemock}
\eeq
and the unitarity of the mock $S$-matrix, one can check that indeed
%$(p_A^{11})^2=p_A^{11}$
\beq
p_A^{11} p_B^{11} =p_A^{11} \delta_{A,B}.
\eeq

We now introduce the projector $Q_1$ onto the fluxons in the $11$ sector of the tube algebra. By definition
\beq
Q_1 = \sum_{A} p_{A}^{11}, 
\label{eq:q1}
\eeq
where $p_{A}^{11}$ only exists if $A$ is a fluxon, i.e., if $n_{A,1}=1$. The above sum puts a weight $1$ on fluxons and $0$ on non-fluxons, so that \beq 
\tr \, Q_1 = \sum_A \tr \, p_A^{11}= \sum_A n_{A,1}=N_\mathcal{C},
\eeq
which shows that $Q_1$ is closely related to the vector $\mathbf{n}_1$. 

From (\ref{eq:c3}) with $i=1$, one gets
\beq
\sum_{A} p_{A}^{11} = Q^1_{111},
\eeq
which is simply the empty tube. Therefore, the projector $Q_1$ onto the fluxons in the $11$ sector is also the projector $Q^1_{111}$ onto the $11$ sector. 

The action of the modular matrices $S$ and $T$ on the tubes is well-known: they act trivially on the empty tube (see Sec. 28.3 in~\cite{SimonBook}). So that
\beq
S Q_1 = Q_1 \text{ and } T Q_1 = Q_1,
\label{eq:sq1}
\eeq
which are the fluxon identities, Eqs.~(\ref{eq:identityS}) and (\ref{eq:identityT}).

\subsubsection{Noncommutative input category}
In the noncommutative case, we also have that the tube algebra restricted to the $11$ sector is the fusion algebra of the input category and that it decouples from the rest of the tube algebra. However, it is noncommutative and there is no mock-$S$ matrix. A consequence is that the number of tubes $N_\mathcal{C}$ is strictly larger than the vector space dimension $V_{11}$ which is strictly larger than the number of fluxons $N_\mathcal{F}$.

From the $N_\mathcal{C}$ tubes $Q^i_{11i}$, we build $N_\mathcal{C}$ idemnils  $p_{A}^{11,ab}$, corresponding to only $N_\mathcal{F}$ (non-simple) idempotents $P_{A}^{11}=\sum_{a=1}^{n_{A,1}} p_{A}^{11,aa}$ in the $11$ sector (the central idempotents are $P_A = \sum_s P_A^{ss}$). Now $\tr \, P_A^{11} = n_{A,1}$ can be larger than 1. We can again define the projector $Q_1$ onto the fluxons in the $11$ sector by:
\beq
Q_1 = \sum_{A} P_{A}^{11} = \sum_{A,a} p_{A}^{11,aa}.
\eeq

From the transformation (\ref{eq:qmp}) between tubes and idemnils, one has
\beq
Q^1_{111} = \sum_{A,a,b} (M^1_{A,111})_{a,b} \, p_A^{11,ab} = \sum_{A, a}  \, p_A^{11,aa} ,
\eeq
as $(M^1_{A,111})_{a,b}=(\Omega^1_{A,111})_{a,b}=\delta_{a,b}$ (see above Eq.~(40)  in Ref.~\cite{Lin21} or below Eq.~(52) and Eq.~(53) in Ref.~\cite{Lan14}). Therefore $Q_1$ is also the projector onto the $11$ sector, i.e. the empty tube $Q^1_{111}$, from which the fluxon identities (\ref{eq:sq1}) follow as in the previous subsection.

\subsection{Stability inequality}
In order to prove an important stability inequality that $\mathbf{n}_1$ needs to satisfy, we make a detour into anyon condensation. Anyon condensation is a general mechanism that allows one to describe a phase transition from a topological order described by a UMTC $\mathcal{A}$ to another described by a UMTC $\mathcal{U}$~\cite{Bais09}. We therefore reverse the perspective and imagine that, instead of building the Drinfeld center from an input category, we condense some bosons of the Drinfeld center to recover the input category. We will use the general formalism of anyon condensation~\cite{Bais09,Neupert16} and apply it to the particular case in which we start from the Drinfeld center $\mathcal{Z}(\mathcal{C})$ and condense it towards a trivial order. This is known as anyon condensation to the vacuum, which is intimately related to finding how many types of gapped boundaries are possible for a given topological order~\cite{Kitaev12,Lan15,Lan20}. 

Condensation to the vacuum relates the UMTC \mbox{$\mathcal{A}=\mathcal{Z}(\mathcal{C})$} to the trivial UMTC $\mathcal{U}$ (total quantum dimension $\mathcal{D}_\mathcal{U}=1$), via a UFC $\mathcal{T}=\mathcal{C}$, where $\mathcal{U}$ is included in $\mathcal{T}$. Generally speaking, anyon condensation 
\beq
\mathcal{A}\to \mathcal{T} \to \mathcal{U},
\eeq
is described by a rectangular matrix $n_{A,s}$, called restriction or lifting matrix, with $A \in \mathcal{Z}(\mathcal{C})$ and $s \in \mathcal{U}$ (in our case,  $s=1$ only). Here, this matrix is the vector $\mathbf{n}_1$ and the condensing bosons are the fluxons. The condensation equations (see, e.g., Eqs.~(20a) and (20b) in Ref.~\cite{Neupert16}) relate the $S$- and $T$- matrices of the two UMTCs $\mathcal{A}$ and $\mathcal{U}$ as
%
%
%%%%%%%%%%%%%
\beq
T_\mathcal{A}  \mathbf{n}_1=\mathbf{n}_1  T_\mathcal{U} \textnormal{ and } S_\mathcal{A} \mathbf{n}_1=\mathbf{n}_1 S_\mathcal{U}.
\eeq 
%%%%%%%%%%%%%
%
%
As $T_\mathcal{U}=1$ and $S_\mathcal{U}=1$ are trivial $1\times 1$ matrices, and $T_\mathcal{A}=T$ and $S_\mathcal{A}=S$ are the $T$- and $S$- matrices of $\mathcal{Z}(\mathcal{C})$, we find Eq.~(\ref{eq:identityS}-\ref{eq:identityT}). Anyon condensation should also hold in the case where the UFC $\mathcal{T}$ has noncommutative fusion rules, as briefly discussed by Bais and Slingerland~\cite{Bais09}.

In the context of anyon condensation, one requires the commutation of fusion and restriction~\cite{Neupert16,Bais09}, i.e.,
\beq
\sum_C N_{A B}^C n_{C,t} = \sum_{r,s}n_{A,r}n_{B,s} \tilde{N}_{r s}^t .
\label{eq:comm1}
\eeq 
In this appendix (and in the following), in order to avoid confusion, $\tilde{N}$ denotes the fusion matrices for $\mathcal{C}$, whereas $N$ denotes the fusion matrices for $\mathcal{Z}(\mathcal{C})$. The stability inequality follows by taking $t=1$ so that
\beq
\sum_C N_{A B}^C n_{C,1} 
=
%\sum_{r,s}n_{A,r}n_{B,s}N_{r s}^1 =
n_{A,1}n_{B,1} +  \sum_{r\neq 1}n_{A,r}n_{B,\bar{r}} \geqslant n_{A,1}n_{B,1},
\label{eq:stability}
\ee
An advantage of this inequality is that it can serve as a test of the coefficients $n_{A,1}$ without the knowledge of the complete lifting/restriction matrix $n_{A,s}$ and the fusion matrix $\tilde{N}$.
 
An interesting question to ask is: given an achiral topological order characterized by a Drinfeld center $\mathcal{Z}(\mathcal{C})$ with modular matrices $S$ and $T$, what are the possible gapped boundaries or condensations to the vacuum~\cite{Neupert16,Lan15,Lan20} ? 

A partial answer to that question is known (see, e.g., Refs~\cite{Lan15,Kawahigashi15,Lan20}): a necessary (but not sufficient) condition for a gapped boundary is to have a vector $\mathbf{n}_1$ that satisfies the fluxon identities Eqs.~(\ref{eq:identityS})-(\ref{eq:identityT}) and the stability condition (\ref{eq:stability}), i.e., to have a stable fluxon set. This is equivalent to the notion of Lagrangian algebras in $\mathcal{Z}(\mathcal{C})$~\cite{Cong16}. For each such stable $\mathbf{n}_1$, there is a corresponding boundary theory described by a UFC $\mathcal{C}_b$. Obviously, $\mathcal{C}$ is one of the possible boundary theories $\mathcal{C}_b$. Also, all the boundary theories are Morita-equivalent, i.e. $\mathcal{Z}(\mathcal{C}_b)\simeq \mathcal{Z}(\mathcal{C})$. This is the bulk-boundary correspondence for achiral topologically-ordered phases. 

For example, given $\mathcal{Z}(S_3)$ (see Sec.~\ref{sec:vecrep} for the notations), one finds five possible $\mathbf{n}_1$'s satisfying (\ref{eq:identityS}) and (\ref{eq:identityT}), one of which ($\mathbf{n}_1=\{1,1,1,0,0,1,0,0\}^T$ which for short we note $\mathbf{n}_1=(A,B,C,F)$ according to its non-zero elements) does not satisfy the stability condition (\ref{eq:stability})~\cite{Lan20}. Among the four stable solutions, two are related by symmetry and correspond to Vec($S_3$) [i.e. $\mathbf{n}_1=(A, B, 2\times C)$ and $\mathbf{n}_1=(A, B, 2\times F)$], and two are related by symmetry and correspond to Rep($S_3$) [i.e. $\mathbf{n}_1=(A, D, F)$ and $\mathbf{n}_1=(A, C, D)$], accounting for the symmetry between the $C$ and $F$ quasiparticles~\cite{Beigi11}. 

Another example is that of $\mathcal{Z}(\mathcal{H}_3)$ (see Sec.~\ref{sec:haa} for the notations). In this case, we find three stable solutions. Two solutions related by symmetry [i.e. $\mathbf{n}_1=\{1,0,0,0,0,0,0,1,2,0,0,0\}^T=(\mathbf{1}, \pi_1,2\times \pi_2)$ and $\mathbf{n}_1=(\mathbf{1}, \pi_1,2\times \sigma_1)$], due to the symmetry between the $\pi_2$ and $\sigma_1$ quasiparticles, point to the fusion ring $H_6$ and one [i.e. $\mathbf{n}_1=(\mathbf{1}, \pi_1, \pi_2, \sigma_1$)] to the fusion ring $H_4$~\cite{Grossman12}. The fusion ring $H_6$ corresponds to the UFCs $\mathcal{H}_3$ and $\mathcal{H}_2$, whereas $H_4$ refers to the UFC $\mathcal{H}_1$. The two categories $\mathcal{H}_3$ and $\mathcal{H}_2$ have the same fusion rules but different $F$ symbols.

%
%
%%%%%%%%%%%%%%%%%%%
%%%%%%%%%%%%%%%%%%%
\section{Generalized Hamiltonian}
\label{app:Hgen}
%%%%%%%%%%%%%%%%%%%
%%%%%%%%%%%%%%%%%%%
%
%

The Hamiltonian that we consider in the main text, Eq.~(\ref{eq:ham}), assigns the same energy penalty +1  to each fluxon which is not the vacuum. However, we can assign different energy penalties to each fluxon type.   Further, we can also assign a plaquette-dependent energy penalties. More generally, let us define a projector 
$$
B_p^A = \left\{
\begin{array}{ll} 1 &  \mbox{if plaquette $p$ has fluxon excitation $A$,}  \\ 
0 & \mbox{otherwise.} \end{array} \right. 
$$
We can then write a more general Hamiltonian
\beq
H_\text{gen} = \sum_{{\text{plaquettes }} \,  p} \,\, \sum_{A \in {\cal F} }  \,\, E^A_p  \, \, B_p^A, 
\label{eq:hamgen}
\eeq
where $E^A_p$ is the energy cost of having fluxon $A$ on plaquette $p$ and $\cal F$ is the set of fluxons. 
So long as $E^{\bf 1}_p < E^A_p$ for all $A \neq {\bf 1}$ and all $p$, then the ground state is still the state where all plaquettes are in the vacuum state.  The Hamiltonian we consider in the main text corresponds to $E_p^A = 0$ for all $A \neq {\bf 1}$ and $E_p^{\bf 1} = -1$, for all plaquettes $p$.  

It is not hard to construct the projectors $B_p^A$ from the structure of the tube algebra by inserting a projector $P_A^{11}$ inside the plaquette $p$ and then fusing it into the edges of the plaquette. For the case where $n_{A,1} > 1$, in fact there is freedom to generalize Hamiltonian Eq.~(\ref{eq:hamgen}) further.  Since the space of states where a plaquette $p$ has fluxon $A$ is $n_{A,1}$-dimensional, one can add terms to split this added degeneracy.   Thus we can even more generally write
\beq
H_\text{gen2} = \sum_{{\rm{plaquettes }} \,  p} \,\, \sum_{A \in {\cal F} }  \,\, \sum_{a=1}^{n_{A,1}} \,\, E^{A,a}_p  \, \, B_p^{A,a},
 \label{eq:hamgen2}
\eeq
where $E^{A,a}_p$ are a set of coefficients and the operators $B_p^{A,a}$ are obtained by inserting the simple idempotent $p_A^{11,aa}$ inside plaquette $p$ and fusing it into the edges.  Again we assure that the vacuum is the ground state when $E^{\mathbf{1}}_p$ is smaller than any $E^{A,a}_p$ for all $A \neq {\bf 1}$ and $a=1,..., n_{A,1}$.

Calculation of spectral degeneracy can in principle be done analytically for any of these Hamiltonians.   One always starts with the Moore-Seiberg-Banks formula, and sums over all possible fluxon labellings of each plaquette.  

Let us take an example of the category Vec($S_3$) as discussed in Sec.~\ref{sec:vecrep}. There are three fluxon types labeled $A,B,C$.  Here $A$ is the vacuum, which we give energy $E^A=0$.  Then $B$ we give energy $E^B=x$.  Finally $C$ is two-dimensional, i.e. $n_{C,1}=2$, and we call $E^{C,1}=y$ and $E^{C,2}=z$ the two values of $E^{C,a}$ with $a=1,2$. For simplicity here, we assume all plaquettes are identical to each other, although this is not necessary. 

Assume that $x,y,z$ are incommensurate, i.e., all ratios of these values are irrational.   If we consider a total energy 
\be
E = n_x x + n_y y + n_z z,
\ee
with integers $n_x, n_y, n_z,$ then the number of eigenstates of the system having exactly this energy is given by the number of ways we can have $n_x$ plaquettes having $B$ fluxons, and $n_y + n_z$ plaquettes having $C$ fluxons with $n_y$ of these in the $y$ eigenstate and $n_z$ of these in the $z$ eigenstate, and all remaining plaquettes in the vacuum eigenstate.   This number is given by 
\begin{eqnarray}
&&\frac{N_\text{p}!}{n_x! n_y! n_z! (N_\text{p} - n_x - n_y -n_z)!}  \times  \nonumber\\
&&  {\rm dim}_{{\cal Z}(S_3)}(g, n_x \mbox{ fluxons } B, (n_y +  n_z) \mbox{ fluxons } C). \quad \quad \quad 
\end{eqnarray}

%
%
%%%%%%%%%%%%%%%%%%%
%%%%%%%%%%%%%%%%%%%
\section{Hilbert-space dimension}
\label{app:hilbspacedim}
%%%%%%%%%%%%%%%%%%%
%%%%%%%%%%%%%%%%%%%
%
%

In this appendix, we use two different methods to count the total Hilbert-space dimension of string-net model having no vertex defects, but having any possible fluxon excitations. In other words, we are restricting the Hilbert space as discussed in Sec.~\ref{sub:fluxon} and counting the size of this restricted Hilbert space,i.e., all possible excitations of the Hamiltonian Eq.~(\ref{eq:ham}). 

The first method we use consists in counting all possible edge and vertex labellings that do not violate the vertex constraints (see Sec.~\ref{sub:countingedge}). The second method, in Sec.~\ref{sub:fluxonfusion}, is to count all possible labelings of the fluxon states of the plaquettes. 

Both calculations are actually sums over fusion networks. In Sec.~\ref{sub:countingedge}, we are summing over fusion trees of $\cal C$ whereas, in Sec.~\ref{sub:fluxonfusion}, we are summing over fusion trees of $\cal Z(\cal C)$   (see the discussion on the derivation of the Moore-Seiberg-Banks formula in Sec.~\ref{subsec:msb}).  In both cases, the fusion trees are sums over fusion multiplicity matrices.  In Sec.~\ref{sub:countingedge} we are summing fusion matrices of $\cal C$ whereas in Sec.~\ref{sub:fluxonfusion} we are summing fusion matrices of ${\cal Z}({\cal C})$.   These two different fusion multiplicity matrices are related to each other via the commutativity of fusion and restriction~\cite{Bais09,Neupert16}. In particular, we have Eq.~(\ref{eq:comm1}), where $r,s,t \in \cal C$ and $A,B,C \in \cal Z(C)$ and $n_{A,r}$ are lifting coefficients. To clearly distinguish the two sets of fusion matrices we call $\tilde N$ that in the input category $\mathcal{C}$ and $N$ that in the output category $\mathcal{Z}(\mathcal{C})$.

For simplicity,  we consider an (orientable) surface of genus $g$ without boundaries ($b=0$) for which the Euler-Poincar\'e characteristic tells us that 
\beq
2 - 2g = N_\text{p} - N_\text{e}  + N_\text{v}.
\eeq
For a trivalent graph, we also have $N_\text{e} =3 N_\text{v}/2$, so that 
\beq
N_\text{p} = 2 - 2g + N_\text{v}/2.
\label{eq:euler}
\eeq  

\subsection{Counting edge labelings}
\label{sub:countingedge}

\begin{figure}
\includegraphics[width=0.8\columnwidth]{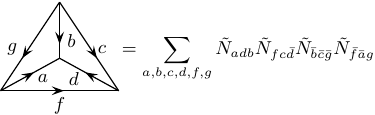}
\caption{Finding the total string-net model Hilbert-space dimension (with no vertex defects) by summing $\tilde N$'s at the vertices over all edge labellings.}  
\label{fig:tetra1}
\end{figure}

Here we want to count up all allowed labellings of the trivalent-graph edges.   Each oriented edge  is labeled  with a simple object in $ \cal C$ and a vertex with incoming arrows $a$ and $b$ and $c$ reading counterclockwise around the vertex  ($a,b,c \in \cal C$) gets a value $\tilde N_{abc} = \tilde N_{ab}^{\bar c}$  (see Fig.~\ref{fig:orientedvertex}). We then sum over all edge labels (summing over $\tilde N$'s is equivalent to counting all vertex labels $\mu \in 1 \ldots \tilde N_{abc}$).  An example of summing over a simple graph is shown in Fig.~\ref{fig:tetra1}.

\begin{figure}[b]
\includegraphics[width=0.6\columnwidth]{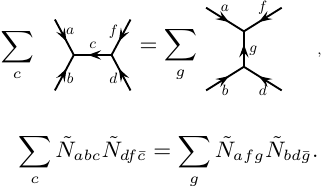}
\caption{The associativity constraint allows restructuring the graph without changing the Hilbert-space dimension.}
\label{fig:fmove2}
\end{figure}

For more complex graphs, one may use restructuring moves (or ``$F$-moves", which here are just the associativity of fusion) to change the connectivity of the graph without altering the Hilbert-space dimension, as shown in Fig.~\ref{fig:fmove2}.

As discussed in Ref.~\cite{Simon13}, if the category has commutative fusion rules, or if the graph is on a surface of genus $g=0$ (a sphere), then by using these restructuring moves, any graph can be reduced to a chain of bubbles 
\beq
\includegraphics{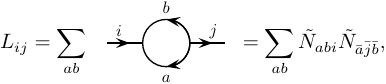}
  \label{dimH_sphere}
\eeq
Thus, for commutative fusion rules, the Hilbert-space dimension is independent of the genus of the surface and only depends on the number of vertices of the graph.

However, in the case where fusion rules are  noncommutative, the situation is different, and on a surface of nonzero genus, one cannot reduce the fusion diagram to a chain of bubbles of the form $L$.  
To understand this, let us first consider the case $g=1$ (torus).    We draw the torus as a rectangle with opposite edges identified.  Using the above restructuring moves (Fig.~\ref{fig:fmove2}), we can reduce any trivalent graph to the following  (where the graph edges are black, and we do not draw arrows or labels on these edges for simplicity). 

\beq
\includegraphics{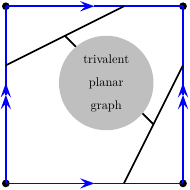}
\label{eq:torpic}
\eeq

If there are multiple lines running around the handles, Eq.~(\ref{eq:torpic}) can be reduced to single lines running around the handle by using restructuring moves (Fig.~\ref{fig:fmove2}). However if there is a single line going around a handle (as shown in the above figure), it cannot be removed by restructuring moves. 

Relaxing the restriction that we need to draw all of our diagrams in the plane, the diagram~(\ref{eq:torpic}) can also be expressed equivalently as
\beq
\includegraphics{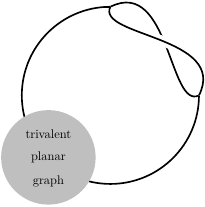}
\eeq

\noindent where the over-crossing could equally well be drawn as an undercrossing, since it is only the connectivity that matters. But note that due to noncommutativity of the fusion rules, it is not possible to remove the under-crossing and flatten the diagram.

We thus define a twisted bubble
\beq
\includegraphics{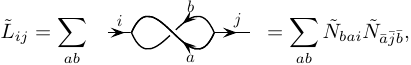}
\label{eq:twistedbubble}
\eeq
where again it does not matter if we drew an over- or under-crossing. 
%since only the connectivity matters here
It is also worth noting that $\tilde L_{1j} = L_{1j}$ since fusion with $1$ is commutative. 

In this notation, the Hilbert-space dimension on a \mbox{$g=1$}-surface (torus) is given by 
\beq
  {\dim \mathcal{H}}  = {\rm Tr}[\tilde L \,\, L^{N_\text{v}/2 - 1}],
\eeq
since two vertices occur within the twisted bubble and the remaining vertices are within the planar graph which can be reduced to a bubble chain of $L$'s.   Note that for the case of commutative fusion we have $\tilde L = L$ and we recover the previous result (\ref{dimH_sphere}).

To extend this to a genus-$g$ surface, we start with the planar representation of this surface as a $4g$-sided polygon with certain edges identified as in the following picture 
\beq
\includegraphics[width=\columnwidth]{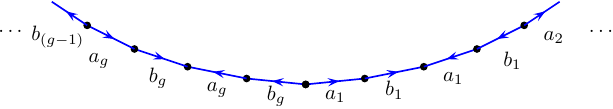}
\eeq

As with the simple torus, a trivalent graph within this polygon can always be reduced, but a single line around a handle cannot be removed. We can thus reduce any graph to the following
\beq
\includegraphics[width=\columnwidth]{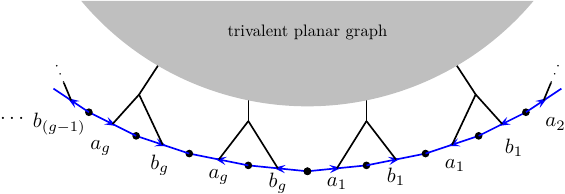}
\eeq

Again, analogous to what we did with the torus, we can remove the periodic boundary conditions and achieve the same connectivity if we allow over- or under-crossings of the graph.  In particular this is entirely equivalent to 
\beq
\includegraphics[width=\columnwidth]{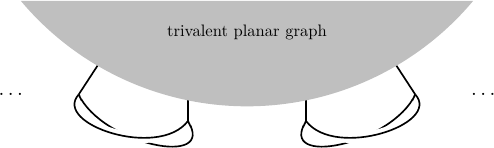}
\label{eq:fodm}
\eeq

So we have $g$ copies of what we called $\tilde L$, each accounting for two vertices, and all the remaining vertices are in some trivalent planar graph. 

Unfortunately, this structure is still somewhat inconvenient, so we use restructuring moves to connect together the legs of each twisted bubble, getting
\beq
\includegraphics[width=\columnwidth]{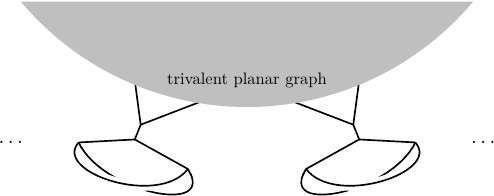}
\eeq

We then define 
\beq
\includegraphics{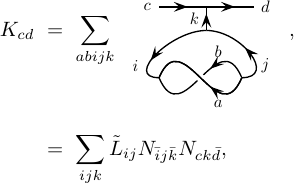}
\eeq
Note that $K$ commutes with $L$ (by associativity of fusion).

Now the total Hilbert-space dimension is then
$$
 {\dim \mathcal{H}} = {\rm Tr}[K^g L^{N_\text{v}/2 - 2 g}],
$$
i.e., there are $g$ twisted loops, each factor of $K$ includes four vertices, and the remaining vertices are in the trivalent planar graph. 

The diagram that gives the twisted bubble $\tilde L$ [see Eq.~(\ref{eq:twistedbubble})] is identical to the diagram we used to define a tube $Q$ in Fig.~\ref{fig:tubeQ}, the difference between the two being only that we have written the tube algebra on a rectangle with sides identified.  $\tilde L_{ij}$  is thus equal to the number of all the tube diagrams $Q^b_{ija}$ summed over $a$ and $b$. In other words, it is the number of tubes $N_\mathcal{T}^{ij}$ in a given $ij$ sector of the tube algebra [see Eq.~(\ref{eq:ntrs})]. Since the tube algebra is also spanned by the idemnils, we can also calculate the  $\tilde L_{ij}$ by counting idemnils $p^{ij,xy}_A$ for all $x,y$ and then summing over all simple objects $A \in \mathcal{Z}(\mathcal{C})$.   Since here $x \in 1 \ldots n_{A,x}$ and $y \in 1 \ldots n_{A,y}$ we have 
\beq
 \tilde L_{ij} =   \sum_A n_{A,i} n_{A,j}.
  \label{eq:cantprove}
\eeq

\subsection{Counting fluxon fusion channels}
\label{sub:fluxonfusion}

We now calculate the same Hilbert-space dimension of a string-net model by summing up all possible fluxon fusion channels assigning all possible fluxons to all  plaquettes. 

We are going to use a graphical notation that a quantum number in $\cal Z(C)$ is a black edge, but a quantum number in $\cal C$ is red.  A lifting coefficient is a green dot. In particular this means that Eq.~(\ref{eq:comm1}) can be expressed graphically as
\beq
\sum_C N_{A B}^C n_{C,t} = \sum_{r,s}n_{A,r}n_{B,s} \tilde{N}_{r s}^t,
\eeq

\beq
\includegraphics{./EqE12.pdf}
\label{eq:graphicid}
\eeq
For simplicity of notation we do not draw arrows on edges. 

To determine the full Hilbert-space dimension, we start with the Moore-Seiberg-Banks formula, which graphically looks like this [see Eq.~(\ref{fig:genusg})]:\\

\centerline{\includegraphics[width=0.45\columnwidth]{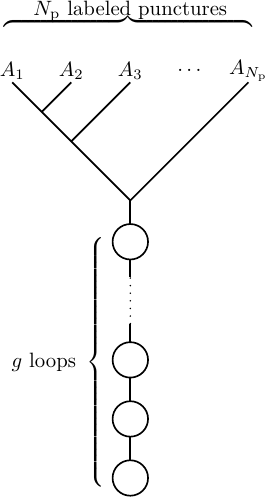}}

Now we want to sum the value of this diagram over the product of $n_{A,1}$ which selects fluxons.  This is  graphically depicted as\\

\centerline{\includegraphics[width=0.45\columnwidth]{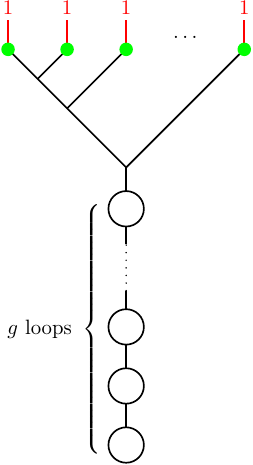}}
 
 Now using the graphical identity Eq.~(\ref{eq:graphicid}) with $t=1$, we can push all the green dots through the vertices to obtain the following:\\

\centerline{\includegraphics[width=0.45\columnwidth]{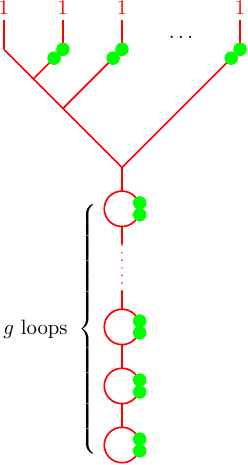}}

Here the double green dot is meant to be connected by a small black line (which is not drawn).  
Given our graphical notation, the double green dot has the following value 
\beq
  \sum_A n_{A,i} n_{A,j} = \tilde L_{ij}  ,
  \label{eq:cantprove2}
  \eeq
where the equivalence to $\tilde L$, the twisted bubble, is from Eq.~(\ref{eq:cantprove}) above.     Note that as mentioned just below Eq.~(\ref{eq:twistedbubble}), we have $\tilde L_{1j} = L_{1j}$ so that the double green dots at the top of the diagram can be replaced by $L$ rather than $\tilde L$.   
Thus the Hilbert-space dimension is given by the diagram

\centerline{\includegraphics[width=0.45\columnwidth]{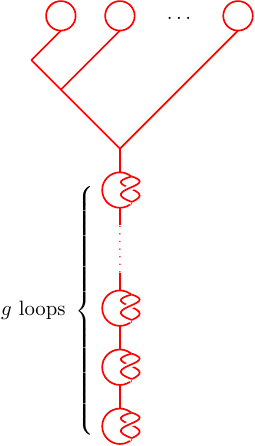}}

This is now a trivalent diagram with edges living in $\cal C$, of the form of Eq.~(\ref{eq:fodm}).  That is, $g$ twisted bubbles connected to a trivalent planar diagram.   

The final step is to check that the number of vertices in this diagram is the same as the number of vertices in the original graph for the string net.   We started with $N_\text{p}$ labeled punctures (at the top of the diagram) and $g$ loops.   The final diagram has $2 (N_\text{p} - 1) - 1$ vertices in the top half of the diagram and $4g -1$ vertices in the bottom half of the diagram, giving a total of $2 N_\text{p} + 4 g - 4$ vertices.   Using Eq.~(\ref{eq:euler}), this is precisely the number of vertices in the original string-net model.

%\bibliography{biblio_Long_SN.bib}

\begin{thebibliography}{60}
\bibitem [{Not()}]{NoteMorita}%
  \BibitemOpen
  \href@noop {} {}\bibinfo {note} {{Actually, the notion of Morita equivalence
  is defined for algebras and not for UFCs. Two algebras that have the same
  modules (i.e. irreducible representations) are said to be Morita equivalent.
  Here if two input categories $\mathcal{C}$ and $\mathcal{C}'$ have tube
  algebras $\mathcal{TA}$ and $\mathcal{TA}'$ that are Morita equivalent, it
  means that they have the same modules. But these are in one to one
  correspondence with the simple objects in the Drinfeld center, so that
  $\mathcal{Z}(\mathcal{C})$ and $\mathcal{Z}(\mathcal{C})'$ are equivalent. By
  abuse of language, we say that two UFCs that have the same Drinfeld center
  are Morita equivalent.}}\BibitemShut {Stop}%
\bibitem [{foo()}]{footnote1}%
  \BibitemOpen
  \href@noop {} {}\bibinfo {note} {{This equation is reminiscent of the
  relation between the central idempotents $P_A$ and the string operators $W_B$
  involving the $S$-matrix [see Eq.~(\ref{eq:PW})]. It relates the projectors
  $p_A^{11}$ onto the fluxons in the $11$ sector to the input string operators
  (or horizontal tubes) $Q^i_{11i}$ via the mock $S$-matrix $\tilde{s}$. It is
  also a particular case of $$ p_A^{11} = \sum _{i \in \mathcal{C}}
  (\mathbf{M}^{-1})_{A,i} \, Q^i_{11i} , $$ which is the inverse of $$
  Q^i_{11i} = \sum_{A\in \mathcal{F}} \mathbf{M}_{i,A} \, p_A^{11}, $$ where we
  introduced the square matrix $\mathbf{M}$ defined by its elements
  $\mathbf{M}_{i,A}=M^i_{A,11i}= \Omega^i_{A,11i} \, d_i$.}}\BibitemShut
  {Stop}%
\end{thebibliography}

%apsrev4-2.bst 2019-01-14 (MD) hand-edited version of apsrev4-1.bst
%Control: key (0)
%Control: author (8) initials jnrlst
%Control: editor formatted (1) identically to author
%Control: production of article title (0) allowed
%Control: page (0) single
%Control: year (1) truncated
%Control: production of eprint (0) enabled
%

\end{document}